\documentclass[12pt,a4paper]{article}

\usepackage[T1]{fontenc}
\usepackage[utf8]{inputenc}
\usepackage[english]{babel}
\usepackage{amsmath,amssymb,mathtools,bm}
\usepackage{graphicx}
\usepackage{wrapfig}
\usepackage{needspace}
\usepackage{subcaption}
\usepackage{float}
\usepackage{placeins}
\usepackage{flafter}
\usepackage{booktabs}
\usepackage{enumitem}
\usepackage{microtype}
\usepackage{setspace}
\usepackage{geometry}
\usepackage{csquotes}
\usepackage{xcolor}
\usepackage[numbers,sort&compress]{natbib}
\usepackage{hyperref}
\usepackage[nameinlink,noabbrev]{cleveref}

\makeatletter
\providecommand{\abx@aux@refcontext}[1]{}
\providecommand{\abx@aux@cite}[2]{}
\providecommand{\abx@aux@segm}[3]{}
\makeatother

\hypersetup{colorlinks=true,linkcolor=blue!40!black,citecolor=blue!40!black,urlcolor=blue!40!black}
\graphicspath{{images/}}

\newcommand{\paperfigure}[2][]{\includegraphics[#1]{#2}}

\newcommand{\dd}{\mathrm{d}}

\newcommand{\bracket}[2]{\left[#1,#2\right]_{\perp}}
\newcommand{\gradperp}{\nabla_{\!\perp}}
\newcommand{\lapperp}{\nabla_{\!\perp}^{2}}

\title{Finite-Larmor-radius vortex interaction from $\delta f$ to full-$f$ gyrofluid dynamics}

\author{Manuel Rinner$^{1}$, F. F. Locker$^{1,*}$, Markus Held$^{2}$, A. Kendl$^{1}$\\[0.5em]
\small $^{1}$Department of Ion Physics and Applied Physics, University of Innsbruck, A-6020 Innsbruck, Austria\\
\small $^{2}$Department of Mathematics and Statistics, UiT The Arctic University of Norway, N-9037 Troms\o, Norway}

\date{}

\begin{document}
\maketitle

\noindent\textit{*Corresponding author:} \href{mailto:franz.locker@uibk.ac.at}{\texttt{franz.locker@uibk.ac.at}}

\begin{abstract}
We study finite-size electrostatic gyrofluid vortex pairs in a two-dimensional isothermal model, focusing on finite-Larmor-radius effects and the transition from local perturbative to finite-amplitude full-distribution dynamics. An effective response kernel based on Padé ion gyroaveraging regularizes the short-range Euler interaction on the Larmor-radius scale while preserving its logarithmic far field, yielding a finite-core prediction for the orbital frequency. Local-model scans show that increasing the ion temperature relative to the electron temperature enhances gyrospinning and filamentation and drives a separation-dependent transition from merger to long-lived co-rotation. Long-wavelength simulations reproduce the predicted finite-Larmor-radius-induced dip in orbital angular frequency, corresponding to a pronounced slowdown of the vortex-pair rotation. For the full-distribution regime, we introduce a quasi-neutrality-consistent initialization that reconstructs the ion gyrocenter density from prescribed electron density and electrostatic potential. It recovers the local limit at small amplitude, whereas finite-amplitude simulations reveal additional mean-density polarization vorticity, delayed or arrested merger, and a loss of density--vorticity sign symmetry already in the cold-ion full-$f$ system. An independent sign scan of density and vorticity then shows how this nonlinear NOB asymmetry combines with the local warm-ion FLR symmetry breaking identified in earlier $\delta f$ vortex studies; for warm ions, Pad\'e-based arbitrary-wavelength NOB and NOB+LWL polarization additionally produce distinct morphologies and trajectories. These results demonstrate that nonlinear density-dependent polarization and FLR physics form separate but coupled mechanisms that can qualitatively alter elementary vortex interactions in two-dimensional gyrofluid dynamics.
\end{abstract}


\section{Introduction}\label{sec:intro}

Isolated vortices are among the most persistent coherent structures of nonlinear fluid motion and provide a particularly transparent route to the elementary dynamics underlying two-dimensional turbulence. In an ideal two-dimensional fluid, kinetic energy and enstrophy are simultaneously conserved, producing the familiar dual-cascade phenomenology in which energy is transferred towards larger scales while enstrophy is transported towards smaller scales \cite{Kraichnan1980,Boffetta2012}. The dynamics of this transfer is not exhausted by spectral arguments: it is realized through the formation, deformation, merger and filamentation of coherent vortices. Numerical and experimental studies have linked the inverse cascade to the elongation and thinning of small vortices by large-scale strain, while filamentary structures are central to the forward transfer of enstrophy \cite{Chen2006,Bruneau2007}. At the level of a vortex pair, mutual rotation, co-advection, merger and the generation of vorticity sheets already reproduce many of the basic nonlinear ingredients of a turbulent vortex population \cite{Saffman1993,Newton2001,Aref2007,Leweke2016}. Such deliberately reduced configurations are therefore valuable not because an isolated vortex pair is expected to occur unmodified in a magnetized plasma, but because they isolate the interaction law that is otherwise embedded in a much more complicated turbulent state.

Magnetized plasmas constitute an important physical realization of quasi-two-dimensional vortex dynamics. For low-frequency fluctuations in a strong magnetic field, the leading perpendicular motion is the $\bm E\times\bm B$ drift, $\bm v_E=\bm E\times\bm B/B^2$, with $\bm E=-\nabla\phi$. In a straight and approximately homogeneous guide field the electrostatic potential consequently plays the role of a stream function, while the corresponding perpendicular flow vorticity is proportional to $\nabla_\perp^2\phi$. This close correspondence between drift dynamics and two-dimensional fluid motion has long motivated reduced descriptions of plasma turbulence and coherent structures \cite{HortonHasegawa1994,Horton1999}. In magnetically confined plasmas the same nonlinear $\bm E\times\bm B$ advection contributes directly to turbulent cross-field particle and heat transport, while self-generated zonal and mean flows can in turn shear and regulate the turbulence \cite{ScottKendlRibeiro2010,KendlScott2003,Lin1998}. 

The neutral-fluid analogy becomes incomplete when perpendicular scales approach the ion gyroradius. A charged ion does not sample the electrostatic potential only at its instantaneous gyrocenter position, but experiences an orbit-averaged field over its Larmor motion. At $k_\perp\rho_i\sim 1$, finite-Larmor-radius (FLR) effects therefore modify both the gyrocenter advection and the polarization relation connecting the particle and gyrocenter densities to the electric potential. Gyrokinetic theory provides the systematic low-frequency reduction in which the fast gyrophase is removed while the relevant orbit-width, kinetic and polarization effects are retained \cite{Hahm1988,BrizardHahm2007,Krommes2012}. Gyrofluid models proceed by taking velocity-space moments of the gyrokinetic distribution and closing the resulting hierarchy, retaining gyroaveraging and polarization physics at a substantially reduced computational cost \cite{Knorr1988,DorlandHammett1993,BeerHammett1996,Scott2003,Scott2005,Scott2010}. In this description the evolved densities are gyrocenter rather than particle densities, and the electrostatic field is obtained from quasi-neutrality together with the polarization response. The underlying role of polarization in guiding-center and gyrocenter reductions has a long theoretical history \cite{Pfirsch1984,Brizard2013}. This distinction is essential here: in the cold-ion, long-wavelength limit the gyrofluid polarization reduces to a Poisson equation and the vorticity dynamics approaches the two-dimensional Euler equation, whereas at finite ion temperature the gyroaveraging operators alter the connection between density, potential and vorticity.

FLR effects on coherent plasma structures have been observed in several related settings. Knorr \textit{et al.} already showed that finite-Larmor-radius corrections can qualitatively alter a vortex-merger calculation \cite{Knorr1988}. In interchange-driven filament dynamics, FLR corrections break poloidal symmetry and generate sheared or spinning flows around the structures \cite{Madsen2011}. For finite-amplitude blobs, ion-temperature effects can alter coherence, morphology and propagation, while nonlinear polarization introduces amplitude-dependent inertia and blob--hole asymmetries \cite{Wiesenberger2014,Kendl2015,Held2016}. More generally, rotational dynamics of plasma blobs can be important for their stability and propagation \cite{Dippolito2004}. These results establish that ion gyro-orbit physics cannot in general be treated as a perturbative correction to otherwise Euler-like motion: it can modify the effective velocity field, the redistribution of vorticity, and the lifetime of coherent structures.

The direct conceptual predecessor of the present work is the gyrofluid vortex-interaction study \cite{vortexmerging}. That work reduced the problem to two-dimensional, isothermal, electrostatic $\delta f$ gyrofluid dynamics and showed explicitly how the classical vortex-pair picture is reorganized by FLR polarization. In the cold-ion limit, co-rotating vortices undergo the expected parity-related Euler-like merger, while counter-rotating vortices co-advect in the familiar dipolar fashion. For warm ions, however, the evolution is no longer determined by the initial vorticity profile alone. The relative distribution of electron and ion gyrocenter densities fixes the FLR contribution to polarization, so two states with identical initial vorticity can evolve differently. Spatial asymmetry between density and vorticity contours generates gyrospinning: spiral arms form around the vortex cores, vorticity is wound into thin sheets, and the resulting structures can either accelerate merger or drive separation depending on the underlying gyrocenter-density configuration \cite{vortexmerging}. The same mechanism also modifies counter-rotating co-advection and was shown to persist in many-vortex drift-wave turbulence modelled with the Hasegawa--Wakatani system, where FLR effects enhance vorticity filamentation and alter the small-scale structure of the flow \cite{vortexmerging,HasegawaWakatani1984}.

That observation is particularly important because vorticity sheets are not merely a visual by-product of vortex deformation. In ordinary two-dimensional turbulence, vortex thinning and filamentation are closely connected with inter-scale transfer \cite{Chen2006,Bruneau2007}; in drift-wave turbulence, nonlinear self-organization and dissipation likewise depend sensitively on the generation of fine vorticity structure \cite{Manz2012}. The FLR-induced sheets in gyrofluid vortices therefore provide a plausible microscopic route by which ion-gyroradius physics can influence the spectral and transport properties of a turbulent plasma. Previous works isolated this mechanism, rather than embedding it immediately in a fully realistic edge model, precisely because the fate of a vortex pair gives a clean measure of how polarization changes nonlinear advection \cite{vortexmerging}. The present paper follows the same philosophy but asks how far that picture survives once two assumptions of the local model are relaxed or examined more closely: the form of the effective interaction at finite $\rho_i$, and the small-amplitude $\delta f$ approximation itself.

The first question is analytical. Classical point-vortex and finite-core theories express the orbital motion of a symmetric vortex pair through the Green function that connects vorticity to the stream function \cite{Saffman1993,Newton2001,Aref2007}. In a gyrofluid plasma, the electric potential plays the role of that stream function, but it is obtained through a scale-dependent polarization response containing gyroaveraging operators rather than through a simple Laplacian inversion. This suggests a natural reformulation: FLR effects may be viewed as a modification of the interaction kernel itself. Such a representation makes it possible to separate the long-range Euler-like part of the interaction from the short-range gyro-orbit correction and to derive an explicit prediction for the orbital frequency of finite-core vortices. It also provides a more quantitative basis for interpreting the dependence on ion temperature and vortex separation than a purely morphological comparison of merger snapshots.

The second question concerns fluctuation amplitude. We denote the electron gyrocenter density by $n_e$, the ion gyrocenter density by $N_i$, and the electrostatic potential by $\phi$. The vortex problem in Ref.~\cite{vortexmerging} is formulated in a local $\delta f$ model, in which the dynamical fields (e.g. species gyrocenter density $\widetilde n_s$) are small perturbations about a prescribed stationary background $N_{\mathrm{bg}}$. This approximation is extremely powerful for turbulence and stability problems satisfying $|\widetilde n_s|/N_{\mathrm{bg}}\ll1$, because the background density can be removed from the nonlinear part of the polarization operator and the field inversion remains comparatively simple. The approximation is not uniformly valid, however, for structures whose density variation is comparable with the reference level. Such amplitudes occur naturally for coherent filaments and blobs in the plasma edge and scrape-off layer, motivating global or full-$f$ gyrofluid descriptions in which the complete moments, rather than only perturbations, are evolved \cite{Dippolito2004,madsenfullf,Wiesenberger2014,Kendl2015,Held2016}. In an unapproximated full-$f$ polarization equation the complete, spatially varying ion gyrocenter density enters the polarization response; in the terminology used below this is the non-Oberbeck--Boussinesq (NOB) formulation. Spatial variations of $N_i$ therefore modify the local polarization inertia and can produce dynamics that have no counterpart in a constant-background $\delta f$ inversion. The passage from $\delta f$ to full-$f$ is consequently not just a change in normalization or bookkeeping; it changes the field equation that generates the $\bm E\times\bm B$ flow.

A further distinction is required between amplitude effects and wavelength effects in the polarization equation. The Oberbeck--Boussinesq (OB) approximation replaces the spatially varying density multiplying the polarization response by a fixed reference value. Thus the unapproximated full-$f$ polarization used here is NOB; applying OB to a full-$f$ model would be an additional approximation. Independently, one may retain the polarization response at arbitrary perpendicular wavelength or expand it in the long-wavelength limit (LWL), $k_\perp\rho_i\ll1$. Historically, full-$f$ gyrofluid polarization was commonly restricted to the LWL because of the short-wavelength closure problem. Held, Wiesenberger and Kendl developed Pad\'e-based full-$f$ closures that retain linear polarization effects at arbitrary wavelength while recovering both the OB and LWL limits \cite{Held2020}, and Held and Wiesenberger later demonstrated explicitly that NOB effects and arbitrary-wavelength effects can alter different aspects of finite-amplitude dynamics \cite{HeldWiesenberger2023}. In the terminology of the present paper, \emph{NOB} therefore denotes full-$f$ polarization without the OB replacement, while \emph{NOB+LWL} denotes the same full-$f$/NOB density dependence combined with the additional long-wavelength approximation. This distinction is especially relevant for vortex interaction because the FLR mechanism of Ref.~\cite{vortexmerging} acts through thin, high-wavenumber vorticity sheets. Recent full-$f$, full-$k$ implementations have extended this framework to quasi-two-dimensional drift-wave and interchange turbulence \cite{KendlTIFF2024}, and three-dimensional full-$f$ gyrofluid turbulence calculations demonstrate the continued development of global gyrofluid modelling beyond simplified two-dimensional benchmarks \cite{WiesenbergerHeld2024}.

A controlled full-$f$ continuation of the vortex-pair problem nevertheless introduces an additional difficulty that is absent in the local calculation: initialization. In the $\delta f$ model, electron and ion gyrocenter perturbations can be chosen so that the linear polarization equation generates a prescribed Gaussian vorticity. In full-$f$, prescribing $n_e$, $N_i$ and $\phi$ in a way that is directly consistent with the corresponding $\delta f$ vortex is nontrivial because the polarization response is nonlinear and variable-coefficient; the three fields therefore cannot in general be prescribed independently while maintaining quasi-neutrality. A meaningful comparison with the local vortex problem requires a construction that preserves the intended electron-density and potential structure and reconstructs the compatible ion gyrocenter density. The GREENY framework provides both $\delta f$ and full-$f$ formulations, together with NOB arbitrary-wavelength and NOB+LWL polarization options, within one numerical implementation \cite{Locker2025}. It therefore permits the same elementary vortex configuration to be followed continuously from the small-amplitude regime into a genuinely finite-amplitude polarization problem.

The literature thus contains two well-developed but only partially connected lines: local gyrofluid studies that isolate how FLR polarization changes elementary vortex interaction \cite{Knorr1988,vortexmerging}, and global/full-$f$ studies that establish the importance of nonlinear polarization, finite inertia and arbitrary-wavelength response for large-amplitude blobs and turbulence \cite{madsenfullf,Wiesenberger2014,Kendl2015,Held2016,Held2020,HeldWiesenberger2023,KendlTIFF2024,WiesenbergerHeld2024}. What has remained largely unexplored is the direct continuation of the same finite-core, freely interacting vortex pair from one regime into the other. Addressing this gap allows the new physics to be attributed cleanly: one can first identify the FLR modification of the local interaction, verify the $\delta f$ limit, and then determine which additional effects arise only when the complete finite-amplitude density participates dynamically in polarization.

The purpose of the present paper is to make this connection systematically. We first formulate an effective FLR-modified interaction kernel for the local gyrofluid polarization response and derive an orbital-frequency prediction for Gaussian finite-core vortices. We then revisit the $\delta f$ co-rotating vortex problem over a broader range of initial separation and ion temperature, including the separation-dependent transition from complete merger to long-lived orbital motion and a comparison of measured orbital frequencies with the analytical prediction. Finally, we construct quasi-neutrality-consistent full-$f$ initial states, verify recovery of the local limit at small amplitude, and determine how nonlinear mean-density polarization modifies the same interaction when the localized density variation becomes comparable with the mean-density level. By varying the density amplitude $n_0$ and vorticity amplitude $\delta$ independently, we then separate the local warm-ion density--vorticity symmetry breaking of Ref.~\cite{vortexmerging} from the additional finite-amplitude sign asymmetry introduced by the full-$f$/NOB polarization operator. The analysis is deliberately restricted to freely evolving electrostatic vortices in a homogeneous guide field. Curvature drive, parallel dynamics, sheath coupling and electromagnetic vortex interaction are excluded so that FLR effects, fluctuation amplitude and the polarization closure can be separated as cleanly as possible. For the summary below, $\tau_i=T_i/T_e$ is the ion-to-electron temperature ratio, $n_0$ is the amplitude of the localized electron-density structure, and $N_{\mathrm{mean}}$ denotes the uniform mean-density level used to construct the full-$f$ initial state; $N_{\mathrm{mean}}$ is not a static background field evolved by the full-$f$ model.

\paragraph{Main results.}
The paper establishes five connected results:
\begin{itemize}[leftmargin=1.6em,itemsep=0.35em,topsep=0.35em]
  \item \textbf{Finite-core analytical interaction law.} A reduced gyrofluid polarization response yields an FLR-modified interaction kernel and an explicit finite-core prediction for the orbital angular frequency of Gaussian vortices.
  \item \textbf{$\delta f$ merger--orbit transition.} In positive-vorticity parameter scans, increasing $\tau_i$ produces a transition from complete merger to a long-lived co-rotating state. The location of this transition in $\tau_i$ depends on the initial vortex separation and is therefore not assigned a universal critical temperature ratio; across the scans, the transition coincides with enhanced small-scale vorticity in the FLR-generated sheets.
  \item \textbf{Recovery and breakdown of the local limit.} A generalized full-$f$ initialization reconstructs the ion gyrocenter density $N_i$ for prescribed electron gyrocenter density $n_e$ and electrostatic potential $\phi$, recovers the $\delta f$ dynamics for sufficiently small localized density amplitudes, and exposes a distinct mean-density polarization contribution when the full-$f$ density variation becomes comparable with $N_{\mathrm{mean}}$.
  \item \textbf{Finite amplitude breaks the local density--vorticity sign symmetry.} An independent scan over $n_0=\pm0.5$ and $\delta=\pm0.01$ shows that full-$f$/NOB polarization distinguishes opposite density departures even at fixed initial vorticity already for cold ions. At finite $\tau_i$, this density-dependent asymmetry combines with the local FLR symmetry breaking known from $\delta f$ vortex dynamics.
  \item \textbf{Arbitrary-wavelength polarization matters once the two mechanisms interact.} For warm ions, the competition between NOB density dependence and FLR-generated fine structure makes the evolution sensitive to whether arbitrary-wavelength NOB or NOB+LWL polarization is retained, producing qualitatively different trajectories and shedding behavior.
\end{itemize}

\FloatBarrier
\section{Gyrofluid model, initialization and diagnostics}\label{sec:model}

\FloatBarrier
\subsection{Electrostatic isothermal gyrofluid equations}\label{sec:equations}

We use the two-dimensional isothermal gyrofluid model implemented in GREENY \cite{Locker2025}, following the full-$f$ polarization structure of global gyrofluid formulations \cite{madsenfullf,Wiesenberger2014}. The geometry is a slab perpendicular to a strong, homogeneous guide field. Electron FLR effects are neglected and the magnetic field magnitude is normalized to unity. In the electrostatic limit, and in the absence of external drives, curvature terms, parallel dynamics and sources, the normalized full-$f$ density equations reduce to
\begin{align}
  \frac{\partial n_e}{\partial t} &= -\bracket{\phi}{n_e},\label{eq:full_ne}\\
  \frac{\partial N_i}{\partial t} &= -\bracket{\psi_i}{N_i},\label{eq:full_Ni}
\end{align}
with the perpendicular Poisson bracket
\begin{equation}
  \bracket{f}{g}=\frac{\partial f}{\partial x}\frac{\partial g}{\partial y}
  -\frac{\partial f}{\partial y}\frac{\partial g}{\partial x}.
\end{equation}
Here $n_e$ is the electron gyrocenter density, $N_i$ the ion gyrocenter density and $\phi$ the electrostatic potential. Throughout this work, gyro-Bohm normalization is employed: densities are normalized to a fixed normalization density $N_0$, perpendicular spatial scales to the drift scale $\rho_S=c_S/\Omega_i$, time to $L_\perp/c_S$, and the electrostatic potential to $T_e/e$, where $c_S=\sqrt{T_e/m_i}$ is the sound speed, $\Omega_i=eB_0/m_i$ the ion gyrofrequency, and $L_\perp$ the perpendicular reference length used in the time normalization. All quantities appearing below are therefore dimensionless, and normalized variables are written without hats. With the normalized ion mass parameter equal to unity, the gyroaveraged ion potential is
\begin{equation}
  \psi_i = \Gamma_1\phi-\frac{1}{2}\left|\gradperp\Gamma_1\phi\right|^2.
  \label{eq:psi_i}
\end{equation}
The Pad\'e FLR operators are
\begin{equation}
  \Gamma_0=\frac{1}{1-\tau_i\lapperp},\qquad
  \Gamma_1=\frac{1}{1-\frac{\tau_i}{2}\lapperp},
  \label{eq:gamma_ops}
\end{equation}
where $\tau_i=T_i/T_e$ controls the ion gyroradius in the adopted gyro-Bohm normalization.

The full arbitrary-wavelength NOB polarization equation is
\begin{equation}
  \gradperp\!\cdot\!\left[
  \sqrt{\Gamma_0}\left(N_i\sqrt{\Gamma_0}\gradperp\phi\right)
  \right]
  =n_e-\Gamma_1N_i.
  \label{eq:full_pol}
\end{equation}
The corresponding NOB long-wavelength-limit (LWL) polarization is
\begin{equation}
  \gradperp\cdot\left(N_i\gradperp\phi\right)=n_e-\Gamma_1N_i.
  \label{eq:lwl_pol}
\end{equation}
Thus, in full-$f$ the ion gyrocenter density enters inside the polarization operator itself. For numerical stabilization, GREENY adds the fourth-order hyperviscous sink $\Lambda_f=-\nu\nabla_\perp^4 f$ to each evolved gyrocenter-density field $f$; the value of $\nu$ used in each simulation set is stated below.

For the local $\delta f$ limit, the normalized prescribed static background is unity and we write
\begin{equation}
  n_e \approx 1+ \widetilde n_e,\qquad N_i \approx 1+\widetilde N_i,
  \qquad |\widetilde n_e|,|\widetilde N_i|\ll1.
\end{equation}
Here $\widetilde n_e$ and $\widetilde N_i$ denote, exclusively, the electron and ion gyrocenter fluctuations about that static $\delta f$ background. We retain the nonlinear perpendicular advection of these fluctuations. The density evolution equations become
\begin{align}
  \frac{\partial\widetilde n_e}{\partial t}&=-\bracket{\phi}{\widetilde n_e},\\
  \frac{\partial\widetilde N_i}{\partial t}&=-\bracket{\Gamma_1\phi}{\widetilde N_i},
\end{align}
closed by the constant-background polarization equation
\begin{equation}
  \frac{1}{\tau_i}(\Gamma_0-1)\phi
  =\widetilde n_e-\Gamma_1\widetilde N_i.
  \label{eq:delta_pol}
\end{equation}
Using $(\Gamma_0-1)/\tau_i=\Gamma_0\lapperp$, the constant-density linearization of \cref{eq:full_pol} reduces directly to \cref{eq:delta_pol}, fixing the sign convention consistently between the full-$f$ and local models.
In the cold-ion limit, $\tau_i\rightarrow0$, this reduces to a Poisson equation for $\phi$, and the electrostatic vorticity $\Omega=\lapperp\phi$ obeys the two-dimensional Euler form $\partial_t\Omega=-\bracket{\phi}{\Omega}$.

\FloatBarrier
\subsection{Density--vorticity symmetry from $\delta f$ to full-$f$}\label{sec:symmetry}

The symmetry breaking discussed in Ref.~\cite{vortexmerging} can be recovered directly from the local equations and then distinguished from the finite-amplitude asymmetry of the full-$f$ polarization used here.  These are related but not identical mechanisms.  The local warm-ion effect shows that a prescribed vorticity does not uniquely specify the gyrofluid state, whereas the full-$f$/NOB effect makes a complete finite-amplitude blob--hole inversion itself nonlinear.

In the cold-ion local limit, the vorticity satisfies
\begin{equation}
  D_\phi\Omega=0,
  \qquad
  D_\phi\equiv\partial_t+\bracket{\phi}{\,\cdot\,},
  \qquad
  \Omega=\lapperp\phi .
  \label{eq:euler_symmetry}
\end{equation}
The density perturbations do not enter this equation, so two cold-ion initial states with the same $\Omega$ have the same vorticity evolution.  For warm ions, a long-wavelength expansion of the local gyrofluid equations gives, to leading FLR order \cite{vortexmerging},
\begin{equation}
  D_\phi\!\left(\Omega+\frac{\tau_i}{2}\lapperp\widetilde n_e\right)
  =\frac{\tau_i}{2}\bracket{\Omega}{\widetilde n_e}
  +\mathcal O\!\left(\tau_i^2\nabla_\perp^4\right).
  \label{eq:df_symmetry_breaking}
\end{equation}
The relevant comparison in Ref.~\cite{vortexmerging} keeps the initial vorticity fixed while changing the density/gyrocenter configuration.  In its simplest sign test,
\begin{equation}
  \Omega\mapsto\Omega,
  \qquad
  \widetilde n_e\mapsto-\widetilde n_e,
  \label{eq:df_density_transform}
\end{equation}
with $\widetilde N_i$ adjusted through \cref{eq:delta_pol} so that the prescribed $\Omega$ remains unchanged.  The two initial generalized vorticities are then
\begin{equation}
  W_{+}=\Omega+\frac{\tau_i}{2}\lapperp\widetilde n_e,
  \qquad
  W_{-}=\Omega-\frac{\tau_i}{2}\lapperp\widetilde n_e,
  \label{eq:df_parity_test}
\end{equation}
and the FLR source changes sign,
\begin{equation}
  \frac{\tau_i}{2}\bracket{\Omega}{-\widetilde n_e}
  =-\frac{\tau_i}{2}\bracket{\Omega}{\widetilde n_e}.
\end{equation}
Thus equal initial vorticity does not imply equal warm-ion evolution: the relative electron--ion gyrocenter-density configuration determines the sign of the FLR polarization response and can select accelerated merger or separation.  This is the precise local symmetry breaking identified by Kendl.  It should not be confused with a complete inversion of all perturbations.  If $(\widetilde n_e,\widetilde N_i,\phi)$ are all reversed, then the polarization relation is reversed consistently and one obtains the corresponding globally sign-reversed state (with the usual reversal of the flow direction).  The non-equivalence in \cref{eq:df_density_transform} instead arises because $\Omega$ is deliberately held fixed while the density configuration is changed.

Full-$f$ polarization introduces a second and independent loss of sign symmetry.  Define the full-$f$ polarization charge
\begin{equation}
  \mathcal Q\equiv n_e-\Gamma_1N_i
  =\mathcal P[N_i,\phi],
  \qquad
  \mathcal P[N_i,\phi]
  \equiv\gradperp\!\cdot\!\left[
  \sqrt{\Gamma_0}\left(N_i\sqrt{\Gamma_0}\gradperp\phi\right)
  \right].
  \label{eq:full_q_definition}
\end{equation}
Writing $N_i=N_{\mathrm{mean}}+\Delta N_i$, its arbitrary-wavelength polarization separates exactly into
\begin{align}
  \mathcal Q
  &=N_{\mathrm{mean}}\Gamma_0\lapperp\phi
  +\gradperp\!\cdot\!\left[
  \sqrt{\Gamma_0}\left(\Delta N_i\sqrt{\Gamma_0}\gradperp\phi\right)
  \right]
  \nonumber\\
  &\equiv \mathcal Q_{\mathrm{OB}}+\mathcal Q_{\mathrm{NOB}}.
  \label{eq:full_pol_decomposition}
\end{align}
The full-$f$ comparison is a genuine finite-amplitude blob--hole inversion about the same uniform mean.  The natural candidate transformation is
\begin{equation}
  (\Delta N_i,\,n_e-N_{\mathrm{mean}},\,\phi)
  \mapsto
  (-\Delta N_i,\,-(n_e-N_{\mathrm{mean}}),\,-\phi).
  \label{eq:full_sign_transform}
\end{equation}
Here both density departures change sign and $\phi\mapsto-\phi$ is the potential that a linear polarization equation would produce.  The density side $\mathcal Q=n_e-\Gamma_1N_i$ is odd under this transformation.  On the polarization side, however, the OB term is odd but the NOB term is even, because it is bilinear in $\Delta N_i$ and $\gradperp\phi$.  Hence
\begin{equation}
  \mathcal Q_{\mathrm{OB}}+\mathcal Q_{\mathrm{NOB}}
  \mapsto
  -\mathcal Q_{\mathrm{OB}}+\mathcal Q_{\mathrm{NOB}}
  \neq-\mathcal Q.
  \label{eq:full_parity_test}
\end{equation}
The transformed fields therefore do not satisfy the full nonlinear polarization equation: the potential of the hole is not simply the negative of the potential of the corresponding blob.  This is the full-$f$/NOB blob--hole symmetry-breaking mechanism.  Density enhancements and depletions carry different polarization inertia even before warm-ion gyroaveraging is considered.

The same result is especially transparent in the LWL, where
\begin{equation}
  \mathcal Q=\gradperp\cdot(N_i\gradperp\phi)
  =N_{\mathrm{mean}}\Omega
  +\Delta N_i\,\Omega
  +\gradperp\Delta N_i\cdot\gradperp\phi .
  \label{eq:full_lwl_symmetry}
\end{equation}
The constant-density term is odd, whereas both NOB terms are even under \cref{eq:full_sign_transform}.  They are of second order for $|\Delta N_i|/N_{\mathrm{mean}}\ll1$, which explains why the local sign symmetry is recovered with the $\delta f$/OB limit, but become comparable to the OB term at finite amplitude.  Locally, if density gradients across a vortex core are neglected, $\mathcal Q\simeq N_i\Omega$ and therefore $\Omega\simeq\mathcal Q/N_i$: equal and opposite polarization charges embedded in a density enhancement and depletion need not generate equal and opposite vorticities.

Finally, full-$f$ and FLR effects act together rather than replacing one another.  Using \cref{eq:full_ne,eq:full_Ni,eq:full_q_definition}, the exact polarization-charge evolution may be written
\begin{equation}
  D_\phi\mathcal Q
  =\Gamma_1\bracket{\psi_i}{N_i}
  -\bracket{\phi}{\Gamma_1N_i},
  \label{eq:full_q_evolution}
\end{equation}
or, after inserting $\psi_i=\Gamma_1\phi+\psi_{2i}$ with
$\psi_{2i}=-|\gradperp\Gamma_1\phi|^2/2$,
\begin{align}
  D_\phi\mathcal Q
  ={}&\underbrace{\Gamma_1\bracket{\Gamma_1\phi}{N_i}
  -\bracket{\phi}{\Gamma_1N_i}}_{\text{finite-$\rho_i$ mismatch of particle and gyrocenter advection}}
  \nonumber\\
  &\underbrace{+\Gamma_1\bracket{\psi_{2i}}{N_i}}_{\text{nonlinear full-$f$ gyrofluid-potential contribution}}.
  \label{eq:full_flr_symmetry}
\end{align}
Equations~\eqref{eq:full_pol_decomposition} and \eqref{eq:full_flr_symmetry} display the combined mechanism.  Under the candidate blob--hole transformation \cref{eq:full_sign_transform}, the two gyroaveraging-mismatch brackets in the first line of \cref{eq:full_flr_symmetry} are even: both $\phi$ and the non-constant part of $N_i$ change sign.  In contrast, $\psi_{2i}$ is quadratic and therefore even in $\phi$, so its bracket with the sign-reversed density departure is odd.  The full evolution consequently contains contributions of different parity in addition to the already non-odd NOB field inversion.  No single transformation $\phi_{\mathrm{hole}}=-\phi_{\mathrm{blob}}$ can map the finite-amplitude warm-ion system onto itself.

In the local small-amplitude limit, $N_i$ is replaced by $N_{\mathrm{mean}}$ inside the polarization operator, the even NOB contribution in \cref{eq:full_pol_decomposition} is discarded, and the higher-order $\psi_{2i}$ coupling is absent from the retained $\delta f$ equations.  In the further cold-ion limit, $\Gamma_1\to1$ and \cref{eq:euler_symmetry} is recovered.  At finite amplitude and finite $\tau_i$, neither reduction applies.  The NOB blob--hole asymmetry of the field inversion and the FLR sensitivity to the relative density--vorticity configuration coexist and jointly determine whether the vortices merge, orbit or separate. This separation of mechanisms is tested directly in Sec.~\ref{sec:full_symmetry_results} by independently reversing the density amplitude and the prescribed vorticity sign.

\FloatBarrier
\subsection{$\delta f$ co-rotating vortex initialization}\label{sec:df_init}

To connect directly to the established gyrofluid vortex benchmark \cite{vortexmerging}, two equal Gaussian electron-density fluctuations of width $a$ are initialized on the prescribed static $\delta f$ background at a center-to-center separation $R_0$,
\begin{equation}
  \widetilde n_{e,0}=n_0\left[g_1(x,y)+g_2(x,y)\right],
\end{equation}
where $n_0$ is the fluctuation amplitude in the $\delta f$ model, and
\begin{equation}
  g_j=\exp\!\left[-\frac{(x-x_0^{(j)})^2+(y-y_0)^2}{a^2}\right],
  \qquad x_0^{(1,2)}=\frac{L_x}{2}\mp\frac{R_0}{2}.
\end{equation}
Following the vortex-interaction construction of Ref.~\cite{vortexmerging}, the ion gyrocenter fluctuation is chosen as
\begin{equation}
  \widetilde N_{i,0}=\Gamma_1^{-1}\left(1-\frac{\delta}{n_0}\Gamma_0\right)\widetilde n_{e,0},
  \label{eq:df_Ni_init}
\end{equation}
where the sign of $\delta$ specifies the initial electron--ion gyrocenter asymmetry. Inserting \cref{eq:df_Ni_init} into \cref{eq:delta_pol} gives the desired Gaussian vorticity pair,
\begin{equation}
  \Omega_0=\lapperp\phi_0=\frac{\delta}{n_0}\widetilde n_{e,0}.
  \label{eq:gaussian_omega_pair}
\end{equation}
For $n_0>0$, the sign of $\delta$ therefore also fixes the sign of the initialized vorticity. At finite $\tau_i$, $\delta$ additionally selects the relative electron--ion gyrocenter-density configuration that drives the FLR gyrospinning response: the positive-$\delta$ branch of the benchmark in Ref.~\cite{vortexmerging} produces an attractive FLR contribution and promotes merger of the co-rotating pair, whereas reversing $\delta$ reverses the initialized gyrocenter asymmetry and selects the branch in which the FLR contribution drives the two cores apart. Thus $\delta$ is not merely an amplitude parameter; it selects the sign and the warm-ion interaction branch of the initialized vortex pair.

\FloatBarrier
\subsection{Quasi-neutrality-consistent full-$f$ initialization}\label{sec:ff_init}

In full-$f$, prescribing $n_e$, $N_i$ and $\phi$ in a $\delta f$-consistent way is nontrivial because of the more complex nonlinear polarization response in \cref{eq:full_pol}. We therefore prescribe the physically desired electron density and electrostatic potential and reconstruct the ion gyrocenter density numerically. To make the comparison with the local initialization explicit, we write the initial electron density as
\begin{equation}
  n_{e,0}=N_{\mathrm{mean}}+\Delta n_{e,0},\qquad
  \Delta n_{e,0}=n_0\left[g_1(x,y)+g_2(x,y)\right],
  \label{eq:full_ne_init}
\end{equation}
where $N_{\mathrm{mean}}$ is the uniform mean-density level of the initial full-$f$ state away from the localized vortex cores and $\Delta n_{e,0}$ is the finite-amplitude density excess relative to that level. Importantly, $N_{\mathrm{mean}}$ is not a prescribed static background field of the full-$f$ dynamics: $n_e$ and $N_i$ remain the complete evolved densities. The ratio $n_0/N_{\mathrm{mean}}$ is used only to characterize the amplitude of the localized structure relative to this mean-density level. 
Introducing \(\phi_G\equiv\sqrt{\Gamma_0}\phi\), we rewrite the full-\(f\) polarization equation by decomposing the complete densities into the constant mean-density level and their finite-amplitude departures, \(N_i=N_{\mathrm{mean}}+(N_i-N_{\mathrm{mean}})\) and \(n_e=N_{\mathrm{mean}}+(n_e-N_{\mathrm{mean}})\). This decomposition is exact and remains valid for finite-amplitude departures. Since the gyroaveraging operators act as the identity on the spatially constant \(N_{\mathrm{mean}}\), the constant terms on the right-hand side cancel, while the mean-density contribution inside the polarization flux yields \(N_{\mathrm{mean}}\nabla_\perp^2\phi_G\). The polarization equation can therefore be written as

\begin{align}
&\nabla_\perp\cdot\left[(N_i-N_{\mathrm{mean}})\nabla_\perp\phi_G\right]
+\sqrt{\Gamma_0}^{-1}\Gamma_1(N_i-N_{\mathrm{mean}})
\nonumber\\
&\qquad =
\sqrt{\Gamma_0}^{-1}(n_e-N_{\mathrm{mean}})
-N_{\mathrm{mean}}\nabla_\perp^2\phi_G .
\label{eq:ff_init_equation}
\end{align}

Thus, the inversion is formulated in terms of the finite-amplitude density departures from \(N_{\mathrm{mean}}\), while the polarization associated with the mean-density level is retained explicitly. The formulation therefore naturally accommodates \(|n_e-N_{\mathrm{mean}}|/N_{\mathrm{mean}}=\mathcal O(1)\), as in the finite-amplitude cases considered below.

We impose homogeneous Dirichlet conditions for the ion-density departure from the mean level,
\begin{equation}
  (N_i-N_{\mathrm{mean}})|_{\partial A}=0.
\end{equation}
The localized Gaussian fields used here decay sufficiently strongly near the domain boundary that the spectral application of the gyro-operators remains compatible with the periodic continuation to numerical accuracy.

Equation \eqref{eq:ff_init_equation} is discretized on a uniform Cartesian grid with a conservative finite-volume treatment of the nonlinear flux. Face-centered gradients of $\phi_G$ define the flux direction and a donor-cell value of $N_i-N_{\mathrm{mean}}$ is used at each face. The FLR and polarization operators are chosen consistently such that the corresponding square-root approximation \(\sqrt{\Gamma_0}\simeq\Gamma_1\) is fulfilled within the adopted second-order Padé representation \cite{Held2020}. Consequently, \(\sqrt{\Gamma_0}^{-1}\Gamma_1\) reduces to the identity contribution in the sparse operator.
 The resulting linear system, where $p$ indexes the set $\mathcal I$ of interior grid cells,
\begin{equation}
  \mathbf M\bm x=\bm b,\qquad x_p=(N_i-N_{\mathrm{mean}})_p,\quad p\in\mathcal I,
  \label{eq:sparse}
\end{equation}
contains at most five non-zero entries per interior row and is solved with BiCGSTAB preconditioned by an incomplete LU factorization. A Laplacian regularization $\epsilon\lapperp(N_i-N_{\mathrm{mean}})$ is available for poorly coupled closed-contour configurations, but was not required for the vortex simulations discussed below. The standard inversion tolerance was $10^{-8}$ with a maximum of 500 iterations.

\begin{wrapfigure}{r}[0pt]{0.42\textwidth}
  \centering
  \captionsetup{font=small}
  \vspace{-0.45\baselineskip}
  \begin{subfigure}[t]{1.16\linewidth}
    \centering
    \paperfigure[width=\linewidth,height=0.23\textheight,keepaspectratio]{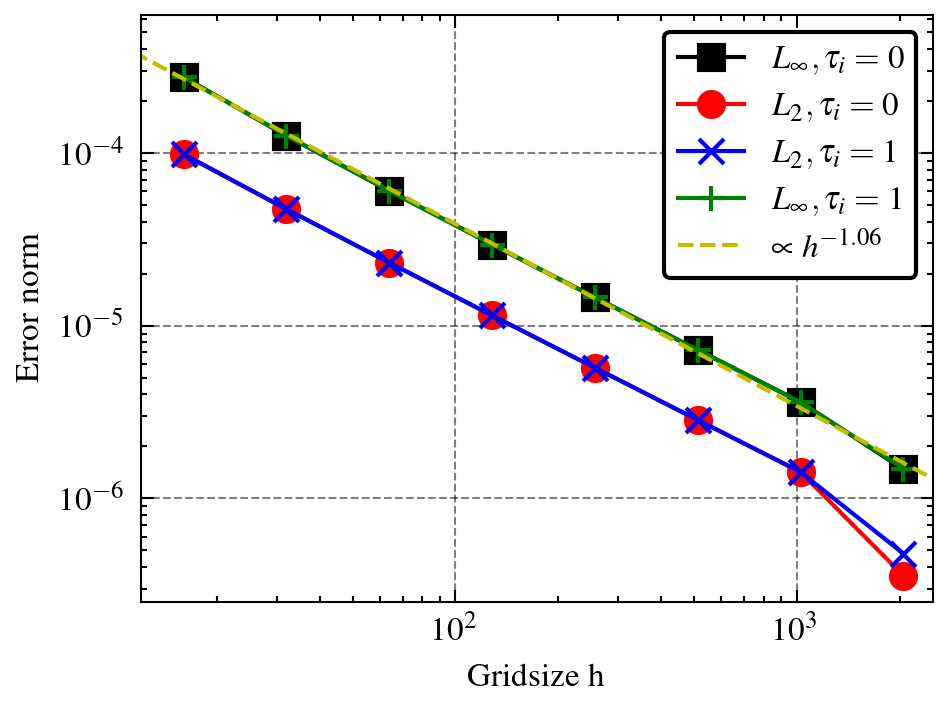}
    \caption{Resolution scaling.}
  \end{subfigure}

  \vspace{0.20em}
  \begin{subfigure}[t]{1.16\linewidth}
    \centering
    \paperfigure[width=\linewidth,height=0.23\textheight,keepaspectratio]{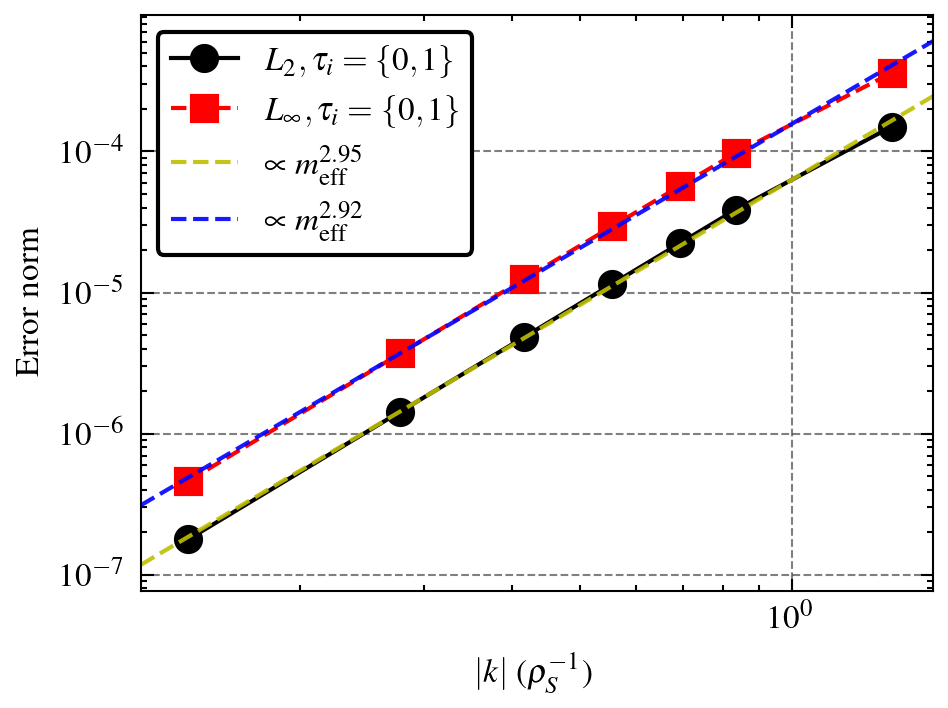}
    \caption{Effective-wavenumber scaling.}
  \end{subfigure}

  \vspace{0.20em}
  \begin{subfigure}[t]{1.16\linewidth}
    \centering
    \paperfigure[width=\linewidth,height=0.23\textheight,keepaspectratio]{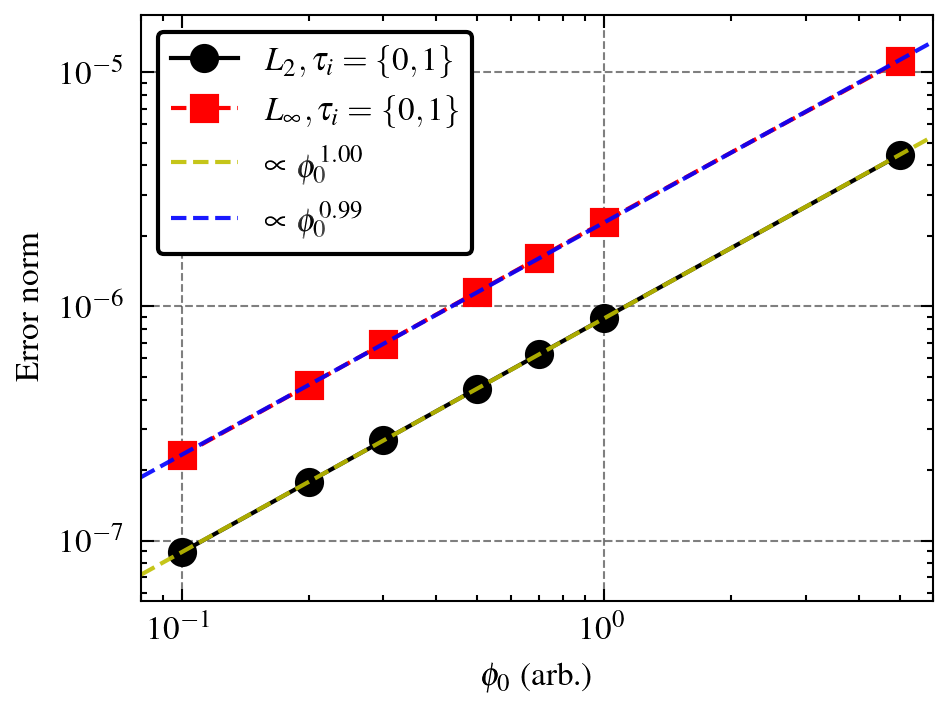}
    \caption{Potential-amplitude scaling.}
  \end{subfigure}
  \caption{Verification of the generalized full-$f$ initialization inversion. The error norms show approximately first-order convergence with grid resolution, cubic scaling with effective wavenumber, and linear scaling with potential amplitude, as expected for the employed discretization.}
  \label{fig:ff_verification}
  \vspace{-0.25\baselineskip}
\end{wrapfigure}

The initialization is verified by manufactured solutions and by recovery of the $\delta f$ limit. For the numerical test we prescribe
\begin{align}
  \phi_G(x,y)&=A\sin^2(k_xx)\sin^2(k_yy),\\
  N_{i,\mathrm{exact}}(x,y)&=N_{\mathrm{mean}}+B\sin^2(k_xx)\sin^2(k_yy),
\end{align}

construct the corresponding $n_e$ analytically from \cref{eq:ff_init_equation}, and compare the recovered $N_i$ with the exact field. The first-order donor-cell discretization yields the expected approximately first-order convergence in grid resolution; the errors also reproduce the expected scaling with effective mode number and potential amplitude. The three verification measures are collected in \cref{fig:ff_verification}.

\Needspace{0.80\textheight}
\subsection{Numerical diagnostics}\label{sec:diagnostics}

The vortex positions are obtained from a local center-of-mass tracker, but the tracked electron-density field must be distinguished between the two models. In the $\delta f$ calculations the weight is the evolved fluctuation $\widetilde n_e$ about the prescribed static background. In the full-$f$ calculations there is no corresponding static background or fluctuation variable; instead we use the finite-amplitude density excess
\begin{equation}
  \Delta n_e(\bm r,t)=n_e(\bm r,t)-N_{\mathrm{mean}},
\end{equation}
relative to the mean-density level introduced for the matched initial state. This subtraction is a diagnostic decomposition of the complete full-$f$ density and must not be identified with the $\delta f$ fluctuation $\widetilde n_e$. Denoting the appropriate tracking field by
\begin{equation}
  n_{\mathrm{trk}}=
  \begin{cases}
    \widetilde n_e, & \delta f,\\
    \Delta n_e, & \text{full-}f,
  \end{cases}
\end{equation}
the center of vortex $j$ at output time $t_i$ is computed inside a disk of radius $\rho$ around its previous position,
\begin{equation}
  \bm c_i^{(j)}=
  \frac{\sum_{|\bm r-\bm c_{i-1}^{(j)}|\leq\rho}\bm r\,n_{\mathrm{trk}}(\bm r,t_i)}
       {\sum_{|\bm r-\bm c_{i-1}^{(j)}|\leq\rho}n_{\mathrm{trk}}(\bm r,t_i)}.
\end{equation}
In the tracking procedure used here, the local radius is typically chosen in the range $\rho=5$--$15$ grid cells, with smaller values used when strong deformation or breakup requires tighter localization.
The separation diagnostic is
\begin{equation}
  R(t)=|\bm c^{(2)}(t)-\bm c^{(1)}(t)|.
\end{equation}
Early-time accelerations are extracted from a quadratic fit to $R(t)$, while the orbital angular frequency is obtained from a linear fit to the pair angle
\begin{equation}
  \theta(t)=\arctan\!\left[
  \frac{c_2^{(2)}(t)-c_2^{(1)}(t)}
       {c_1^{(2)}(t)-c_1^{(1)}(t)}\right]
\end{equation}
with a continuous branch chosen over the initial constant-frequency fit interval.

\FloatBarrier
\section{Analytical vortex-interaction model}\label{sec:theory}

\FloatBarrier
\subsection{From interaction energy to orbital frequency}\label{sec:hamiltonian}

For a two-dimensional incompressible flow with stream function $\psi$ and vorticity $\Omega=\nabla^2\psi$, the vortex-pair interaction can be expressed through a Hamiltonian functional \cite{Aref2007,Newton2001,Saffman1993}
\begin{equation}
  H[\Omega]=-\frac12\int\dd^2r\int\dd^2r'\,
  \Omega(\bm r)G(\bm r-\bm r')\Omega(\bm r').
  \label{eq:Hreal}
\end{equation}
For two identical vortices of circulation $\Gamma$ and a symmetric pair Hamiltonian $H(R)$, the non-canonical point-vortex equations give
\begin{equation}
  \omega(R)=\frac{2}{\Gamma R}\frac{\dd H}{\dd R}.
  \label{eq:omega_H}
\end{equation}
To obtain a form applicable to a general linear inversion kernel,
we define the inverse-polarization kernel by
\begin{equation}
    \hat{\psi}(\mathbf{k})
    =
    -\hat{G}(k)\hat{\Omega}(\mathbf{k}),
\end{equation}
so that $\hat{G}(k)=1/k^2$ in the Euler limit. For two identical axisymmetric vortices, Fourier transformation and angular integration yield a Bessel representation of the pair interaction. A direct interaction-energy calculation for two extended cores contains the product $|\hat\Omega_0(k)|^2$. For Gaussian cores this gives a different finite-core factor from the standard velocity-at-the-center result because the interaction energy averages over both extended distributions. The numerical diagnostic used here instead follows the center of one vortex in the velocity field generated by the other. We therefore use the corresponding single-spectrum form,
\begin{equation}
  \omega_{\mathrm{eff}}(R)=\frac{1}{\pi\Gamma R}
  \int_0^\infty\dd k\,k^2G(k)\hat\Omega_0(k)J_1(kR),
  \label{eq:omega_general}
\end{equation}
where $J_1$ is the Bessel function of the first kind. For the Euler kernel this choice recovers the standard Gaussian finite-core law $\omega=\Gamma(1-e^{-R^2/a^2})/(\pi R^2)$ \cite{Saffman1993}. For a Gaussian vortex
\begin{equation}
  \Omega_0(r)=\Omega_0\exp(-r^2/a^2),\qquad
  \Gamma=\pi a^2\Omega_0,
\end{equation}
its normalized spectral form contributes the finite-core factor $\exp(-a^2k^2/4)$.

\FloatBarrier
\subsection{Reduced gyrofluid response kernel}\label{sec:kernel}

\begin{figure}[!htbp]
  \centering
  \paperfigure[width=0.95\linewidth]{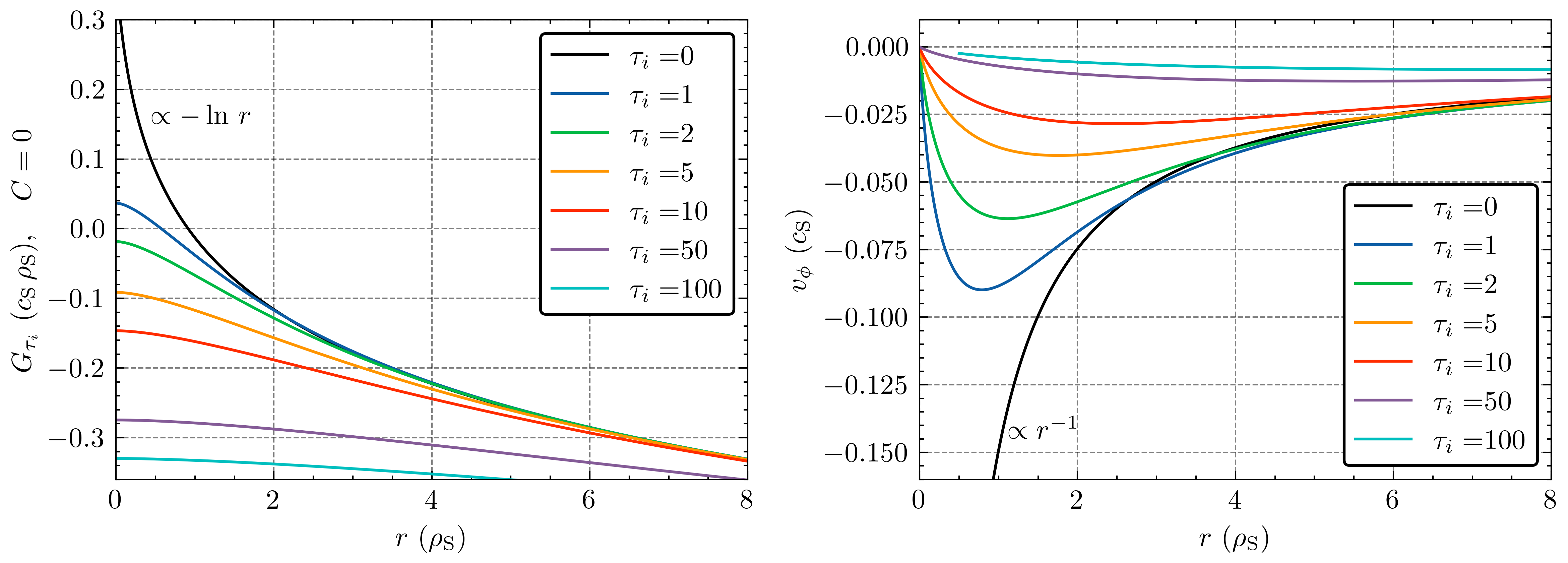}
  \caption{Interaction kernel $G_{\tau_i}(r)$ and radial derivative $\partial_rG_{\tau_i}$ for the reduced $\delta f$ gyrofluid polarization response. In the cold-ion limit the interaction approaches the logarithmic Euler point-vortex potential. Finite $\tau_i$ removes the short-range singularity on the scale $\ell_{\mathrm{FLR}}\sim\sqrt{\tau_i/2}\,\rho_S$ while preserving the asymptotic far-field behavior.}
  \label{fig:interaction_kernel}
\end{figure}

A closed analytical kernel can be obtained from a deliberately reduced electrostatic $\delta f$ polarization response. Starting from \cref{eq:delta_pol}, we retain ion gyroaveraging but replace the potential-side polarization response \((\Gamma_0-1)/\tau_i\) by its long-wavelength form \(\nabla_\perp^2\), thereby neglecting its finite-\(k_\perp\rho_i\) gyroscreening. In Fourier space this gives
\begin{equation}
  -k^2\hat\phi(\bm k)=\hat n_e(\bm k)-
  \frac{1}{1+\frac{\tau_i}{2}k^2}\hat N_i(\bm k).
  \label{eq:reduced_pol}
\end{equation}
Introducing the generalized source
\begin{equation}
  \hat{\widetilde N}=(1+\tfrac{\tau_i}{2}k^2)\hat n_e-\hat N_i,
\end{equation}
we obtain
\begin{equation}
  \hat\Omega=\frac{\hat{\widetilde N}}{1+\frac{\tau_i}{2}k^2}
\end{equation}
and the effective algebraic interaction kernel
\begin{equation}
  \hat G_{\tau_i}(k)=
  \frac{1}{k^2\left(1+\frac{\tau_i}{2}k^2\right)}.
  \label{eq:Gk}
\end{equation}
Using partial fractions, the corresponding real-space kernel is
\begin{equation}
  G_{\tau_i}(r)=
  -\frac{1}{2\pi}\ln r
  -\frac{1}{2\pi}K_0\!\left(\sqrt{\frac{2}{\tau_i}}\,r\right)+C,
  \label{eq:Gr}
\end{equation}
with $K_0$ the modified Bessel function of the second kind \cite[Chapter~10]{NIST:DLMF}. Its radial derivative is
\begin{equation}
  \frac{\dd G_{\tau_i}}{\dd r}=
  -\frac{1}{2\pi r}
  +\frac{1}{2\pi}\sqrt{\frac{2}{\tau_i}}
  K_1\!\left(\sqrt{\frac{2}{\tau_i}}\,r\right).
  \label{eq:dGr}
\end{equation}
The additional Bessel contribution regularizes the $r\rightarrow0$ Euler singularity on the characteristic scale
\begin{equation}
  \ell_{\mathrm{FLR}}=\sqrt{\frac{\tau_i}{2}}\,\rho_S,
\end{equation}
while the far field remains logarithmic and therefore Euler-like, as illustrated in \cref{fig:interaction_kernel}.

For two point vortices, \cref{eq:omega_H,eq:dGr} give
\begin{equation}
  \omega_{\mathrm{point}}(R)=\frac{\Gamma}{\pi R^2}
  \left[1-xK_1(x)\right],\qquad
  x=R\sqrt{\frac{2}{\tau_i}}.
  \label{eq:omega_point}
\end{equation}
For the finite Gaussian cores used in the simulations, \cref{eq:omega_general,eq:Gk} instead give
\begin{equation}
  \boxed{\;
  \omega_{\mathrm{eff}}(R;a,\tau_i)=
  \frac{\Gamma}{\pi R}
  \int_0^\infty
  \frac{J_1(kR)e^{-a^2k^2/4}}
  {1+\frac{\tau_i}{2}k^2}\,\dd k
  \;}
  \label{eq:omega_eff}
\end{equation}
which reduces to the finite-core Euler result for $\tau_i\rightarrow0$ and to the point-pair FLR expression for $a\rightarrow0$. Expanding the polarization response to first non-trivial order at long wavelength gives
\begin{equation}
  \omega_{\mathrm{LWL}}(R;a,\tau_i)=
  \frac{\Gamma}{\pi R}\int_0^\infty\dd k\,
  J_1(kR)e^{-a^2k^2/4}
  \left(1-\frac{\tau_i}{2}k^2\right).
  \label{eq:omega_lwl}
\end{equation}
Numerical evaluation predicts an FLR-dependent dip in the orbital angular frequency. This prediction is tested below.

\FloatBarrier
\section{$\delta f$ vortex interaction}\label{sec:df_results}

The local-model results are presented first in detail. For a direct side-by-side comparison with the finite-amplitude full-$f$ dynamics, the principal findings of both descriptions are summarized at the beginning of \cref{sec:full_results}.

\FloatBarrier
\subsection{Warm-ion parameter scan and arrested merger}\label{sec:arrested}

\begin{figure}[!htbp]
  \centering
  \paperfigure[width=\linewidth]{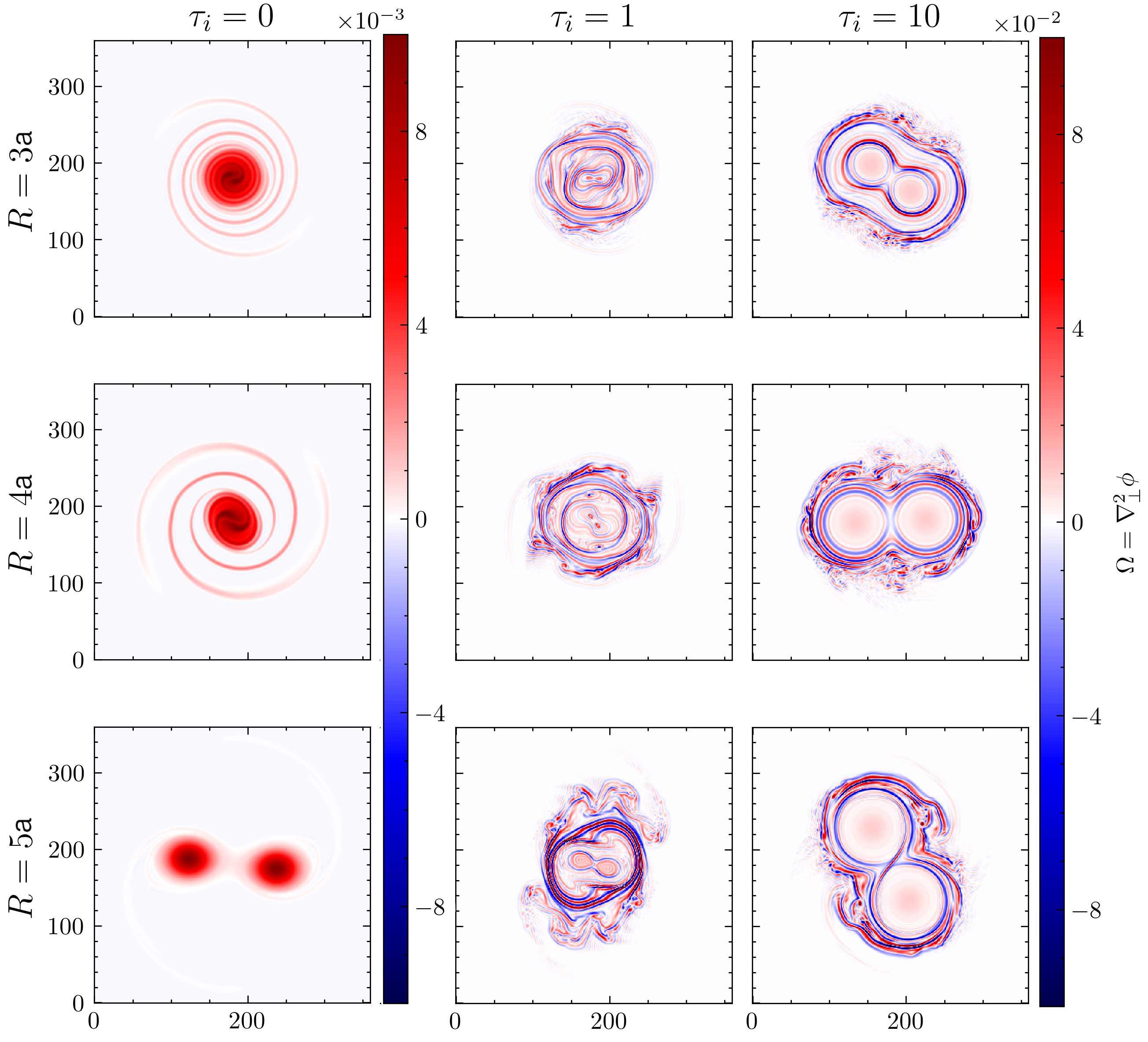}
  \caption{$\delta f$ evolved vorticity for positive initial vorticity amplitude at the common time $t=9000\,L_\perp/c_S$, displayed for different initial separations (rows) and ion-to-electron temperature ratios $\tau_i$ (columns). Cold-ion cases merge fully, whereas the largest-$\tau_i$ cases form a long-lived co-rotating pair at approximately constant separation.}
  \label{fig:df_scan_vorticity}
\end{figure}

\begin{figure}[!htbp]
  \centering
  \paperfigure[width=\linewidth]{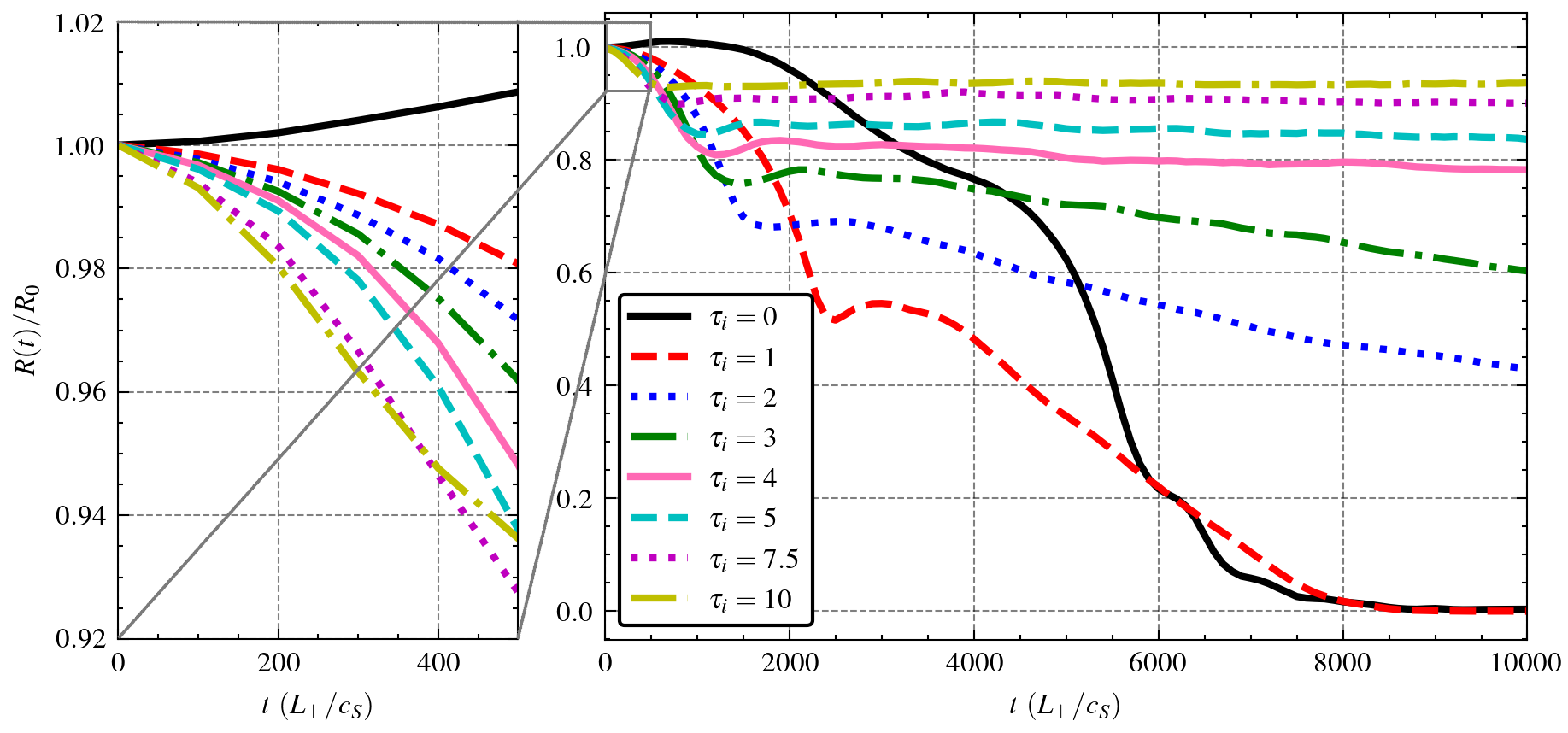}
  \caption{Time evolution of the normalized separation of two positive-vorticity $\delta f$ vortices for different $\tau_i$, including the early-time evolution. Larger $\tau_i$ initially decreases the separation more rapidly, but can also cause the vortices to interlock and remain at an approximately constant distance. The scan shows a transition between merging and long-lived co-rotation, but its location in $\tau_i$ depends on the initial vortex separation and should not be interpreted as a universal threshold.}
  \label{fig:df_scan}
\end{figure}

The $\delta f$ implementation first reproduces the established cold-ion co-rotation and the warm-ion gyrospinning, merger enhancement or repulsion reported in \cite{vortexmerging}. Unless stated otherwise, these $\delta f$ vortex scans use hyperviscosity $\nu=10^{-6}$. We then extend the co-rotating positive-vorticity case over a wider range of separation and $\tau_i$. Representative vorticity fields across the scan are shown in \cref{fig:df_scan_vorticity}, while the corresponding separation histories are summarized in \cref{fig:df_scan}. Cold-ion pairs proceed toward complete coalescence. At finite ion temperature the vortices spin up and develop thin surrounding vorticity sheets. For a given initial separation, sufficiently strong finite-ion-temperature effects can instead prevent complete merger, leaving a long-lived co-rotating pair at approximately constant separation; the required $\tau_i$ shifts with the initial separation.

The extracted arrested-orbit distances increase monotonically with ion temperature over the investigated orbiting cases (\cref{tab:arrested_orbits}). The present data do not justify a unique scaling law for this distance or a universal critical value of $\tau_i$: the location of the merger--orbit crossover shifts with the initial vortex separation. The transition should therefore be interpreted as a numerically observed, separation-dependent change of regime rather than as a fitted critical threshold or exponent.

\begin{table}[t]
  \centering
  \caption{Measured asymptotic separation of the long-lived co-rotating $\delta f$ vortex pairs.}
  \label{tab:arrested_orbits}
  \begin{tabular}{cc}
    \toprule
    $\tau_i$ & orbiting separation $R_{\rm orb}/a$\\
    \midrule
    4   & 3.2\\
    5   & 3.4\\
    7.5 & 3.6\\
    10  & 3.8\\
    \bottomrule
  \end{tabular}
\end{table}

The early-time cold-ion acceleration shows two additional empirical trends in the investigated parameter range: an approximate $R^{-6}$ dependence on initial separation and an approximately quadratic dependence on the initial vorticity amplitude. At finite $\tau_i$ the amplitude dependence becomes substantially more complicated. We therefore do not use these empirical fits as a theoretical scaling law; their main role is to demonstrate that warm-ion interaction cannot be represented by a simple multiplicative correction to the cold-fluid pair dynamics.

\FloatBarrier
\subsection{Small-scale vorticity and the merger transition}\label{sec:enstrophy}

\begin{figure}[!htbp]
  \centering
  \paperfigure[width=0.95\linewidth]{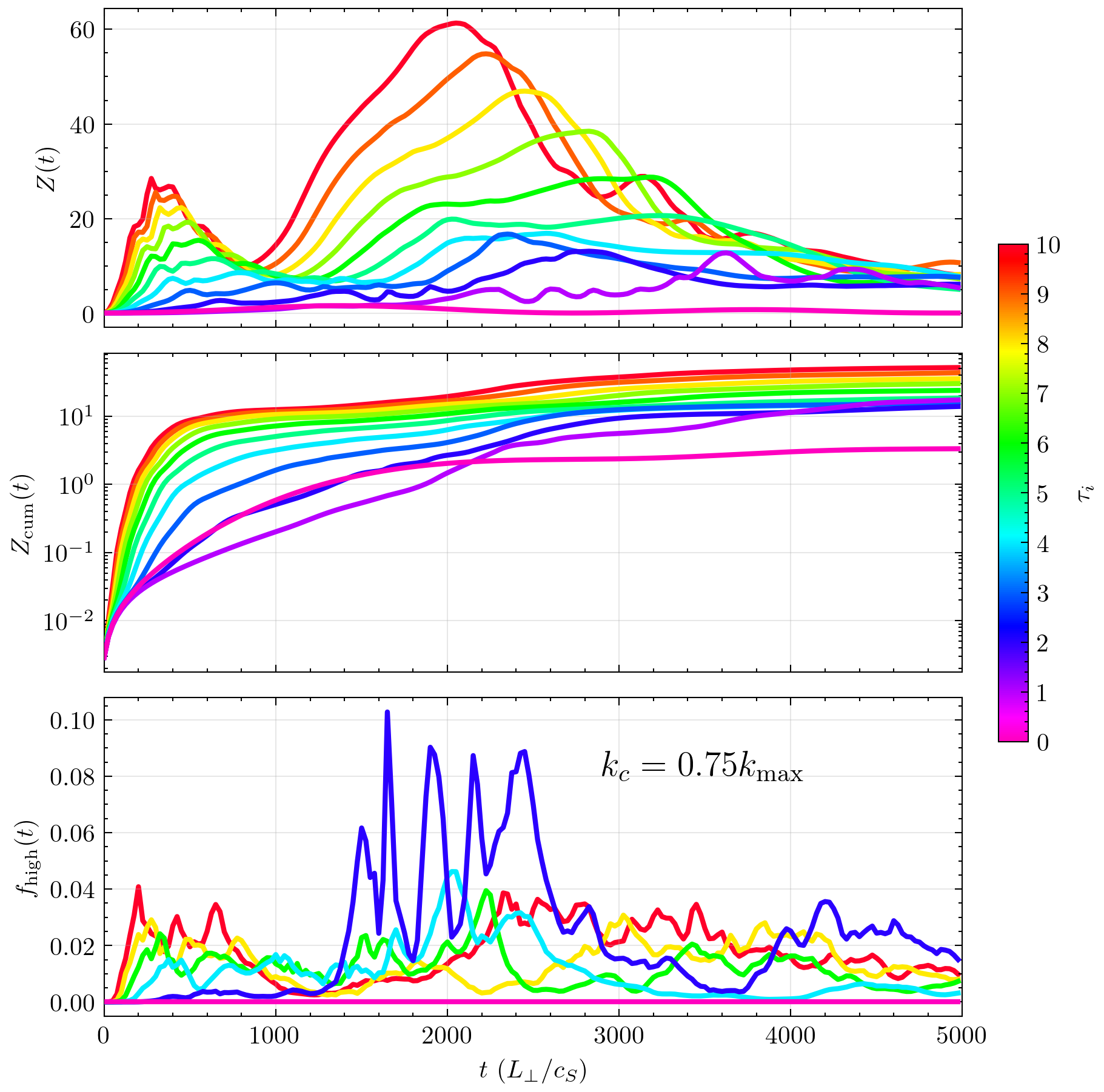}
  \caption{Time evolution of total enstrophy, cumulative dissipative enstrophy loss and the fraction of enstrophy above $k_c=0.75k_{\max}$ for positive-vorticity $\delta f$ vortex pairs at different $\tau_i$. The enhanced high-wavenumber content at larger ion temperature accompanies the transition to long-lived co-rotation.}
  \label{fig:enstrophy}
\end{figure}

The change from merger to arrested co-rotation is accompanied by a redistribution of vorticity toward finer spatial scales. We quantify this through the total enstrophy, the cumulative dissipative enstrophy loss and the fraction of enstrophy above $k_c=0.75k_{\max}$. The total enstrophy is defined as
\begin{equation}
    Z(t)=\frac{1}{2}\int_A dA\,\Omega^2.
\end{equation}
To quantify the enstrophy removed by hyperviscous dissipation, we use the
cumulative dissipative enstrophy
\begin{equation}
    Z_{\rm cum}(t)
    =
    \int_0^t dt'\,
    \nu\int_A dA\,(\nabla_\perp^2\Omega)^2 .
\end{equation}
The fraction of enstrophy contained at high perpendicular wavenumbers is
defined as
\begin{equation}
    f_{\rm high}
    =
    \frac{
        \displaystyle\int_{k_c}^{\infty} dk\,|\hat{\Omega}(k)|^2
    }{
        \displaystyle\int_{0}^{\infty} dk\,|\hat{\Omega}(k)|^2
    },
    \qquad
    k_c=0.75\,k_{\max}.
\end{equation}
As shown in \cref{fig:enstrophy}, larger $\tau_i$ generates stronger early FLR spin-up and a larger high-$k$ fraction. Within a fixed-separation scan, the marked increase in fine-scale enstrophy coincides with entry into the arrested-orbit regime; the corresponding value of $\tau_i$ is separation dependent. On the merging side of the crossover, subsequent density-core interaction continues to generate vorticity during merger. On the long-lived co-rotating side, the interaction is increasingly dominated by the encounter of the enveloping vorticity sheets, while later enstrophy production becomes weak. This supports the interpretation that sufficiently developed FLR-generated sheets inhibit further mixing of the density cores and thereby arrest complete coalescence.

\FloatBarrier
\subsection{Orbital frequency compared with the analytical prediction}\label{sec:omega_comparison}

\begin{figure}[!htbp]
  \centering
  \paperfigure[width=0.95\linewidth]{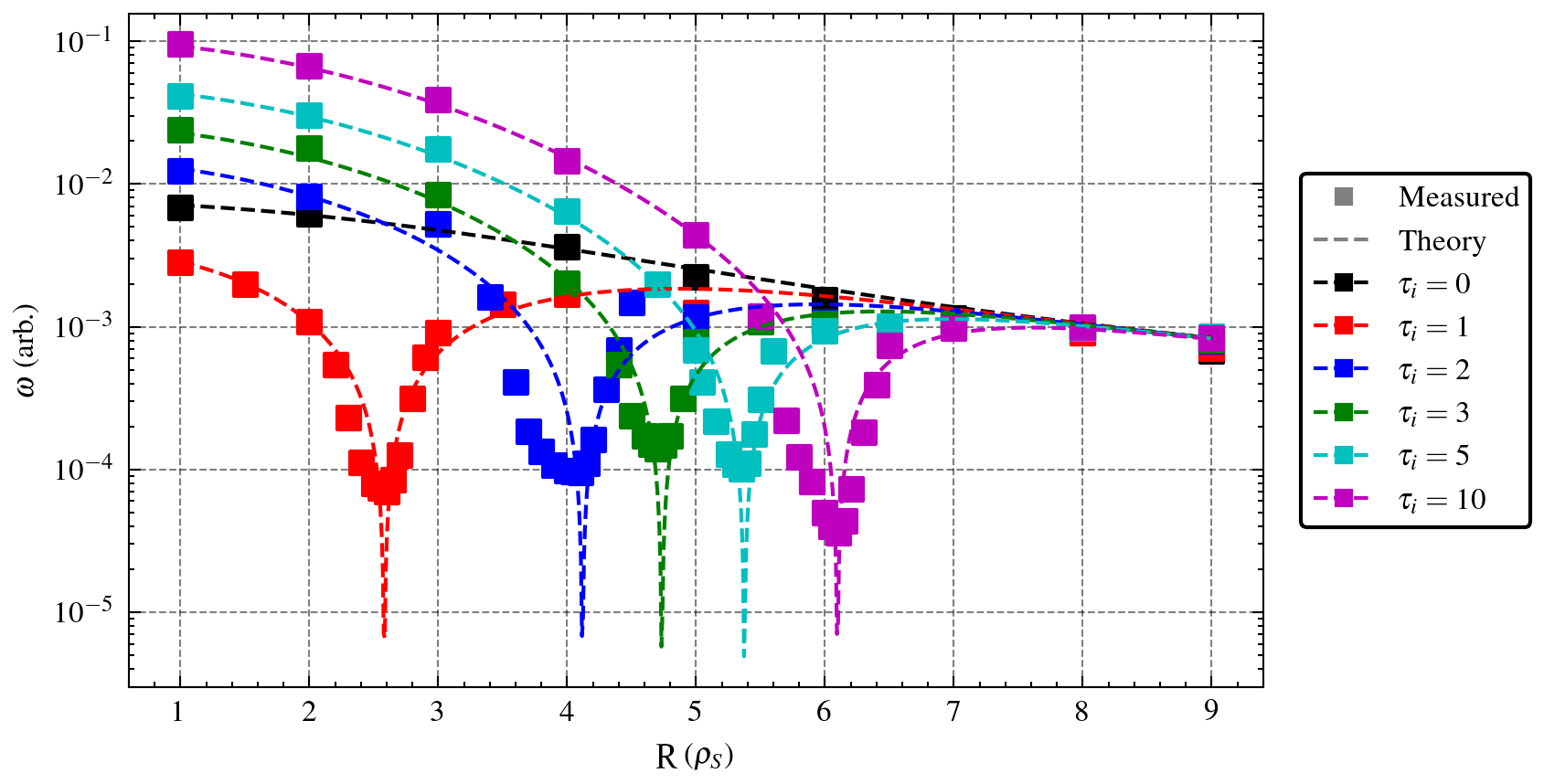}
  \caption{Orbital angular frequency as a function of initial separation $R$ for several $\tau_i$, compared with the long-wavelength finite-core prediction \cref{eq:omega_lwl}. The simulation data were obtained using the $\delta f$ model with LWL polarization and $(L_x,L_y)=(32\rho_S,32\rho_S)$, a numerical resolution of $1024^2$, $a=2\rho_S$ and $\Omega_0=0.01$.}
  \label{fig:angular_dip}
\end{figure}

The finite-core theory of \cref{sec:theory} predicts an FLR-dependent modification of the orbital frequency. The comparison shown in \cref{fig:angular_dip} uses direct $\delta f$ long-wavelength-limit (LWL) simulations in a $(32\rho_S)^2$ domain with $1024^2$ grid points, Gaussian vortex width $a=2\rho_S$ and initial vorticity amplitude $\Omega_0=0.01$. The vortex-center frequency is extracted directly from these LWL runs, without any spectral filtering or post-processing modification of the polarization response. The comparison therefore provides a direct numerical test of the long-wavelength analytical prediction for the same parameter set.

Within this qualification, the simulated orbital frequencies follow the theoretical curves well over the scanned initial separations and reproduce the predicted $\tau_i$-dependent dip. The agreement supports the main physical premise of the analytical model: FLR physics can be interpreted, at the level of the reduced vortex-pair problem, as a modification of the response kernel that connects the generalized source to the advecting electrostatic potential.

\FloatBarrier
\section{From $\delta f$ to finite-amplitude full-$f$ interaction}\label{sec:full_results}

\paragraph{Key findings: $\delta f$ versus full-$f$.}
The principal numerical results can be read most directly by comparing the two descriptions point by point:

\begin{itemize}[leftmargin=1.6em,itemsep=0.55em,topsep=0.4em]
  \item \textbf{Polarization regime.}
  \textbf{$\delta f$:} the background is fixed in the local polarization response and the approximation is controlled for small fluctuations.
  \textbf{full-$f$:} the complete gyrocenter density enters the nonlinear polarization operator, so the complete density dependence participates dynamically once the localized amplitude reaches $n_0/N_{\mathrm{mean}}=\mathcal O(1)$.

  \item \textbf{Small-amplitude limit.}
  \textbf{$\delta f$:} provides the reference local dynamics.
  \textbf{full-$f$:} the generalized initialization converges toward the $\delta f$ separation histories as $n_0/N_{\mathrm{mean}}\rightarrow0$; in the investigated cases the trajectories are already close for $n_0/N_{\mathrm{mean}}=0.01$, while warm-ion deviations remain visible at $n_0/N_{\mathrm{mean}}=0.1$.

  \item \textbf{FLR-mediated interaction.}
  \textbf{$\delta f$:} FLR modifies the effective interaction kernel, drives gyrospinning and generates thin enveloping vorticity sheets. Positive-vorticity pairs change from complete merger to long-lived co-rotation as ion temperature is increased, but the crossover value of $\tau_i$ shifts with the initial vortex separation. Across the orbiting cases in the present scan, $R_{\rm orb}/a$ increases with $\tau_i$.
  \textbf{full-$f$:} the same FLR dynamics acts together with an additional finite-amplitude mean-density polarization contribution, which can delay or arrest the ordinary merger and qualitatively alter the subsequent interaction.

  \item \textbf{Density--vorticity sign symmetry.}
  \textbf{$\delta f$:} at $\tau_i=0$, $D_\phi\Omega=0$ and reversal of the density perturbation at fixed $\Omega_0$ is a symmetry of the vorticity dynamics; at finite $\tau_i$, the FLR source $\propto\tau_i\bracket{\Omega}{\widetilde n_e}$ changes sign and breaks this equivalence, as established in Ref.~\cite{vortexmerging}.
  \textbf{full-$f$:} the NOB contribution is bilinear in the density departure and electric field, so the same density reversal is already non-equivalent at $\tau_i=0$. Warm ions therefore combine two distinct parity-breaking mechanisms: nonlinear NOB polarization and local FLR density--vorticity coupling.

  \item \textbf{Small-scale structure and wavelength response.}
  \textbf{$\delta f$:} increasing $\tau_i$ enhances the high-$k$ enstrophy fraction, and the onset of strongly developed FLR sheets coincides with the merger--orbit transition; direct LWL simulations reproduce the predicted orbital-frequency dip.
  \textbf{full-$f$:} mean-density polarization adds shielded opposite-sign vorticity. For $\tau_i=1$, arbitrary-wavelength NOB polarization retains finer sheets and produces an extended orbit followed by rapid repulsion, whereas NOB+LWL develops coarser structures and vortex shedding for positive initial vorticity; corresponding shedding is not observed in the arbitrary-wavelength NOB case over the simulated interval.

  \item \textbf{Physical interpretation.}
  \textbf{$\delta f$:} finite ion temperature changes how one vortex advects the other through an FLR-modified response kernel.
  \textbf{full-$f$:} finite amplitude changes the vorticity field that mediates the interaction itself through the complete density-dependent polarization response; the global dynamics therefore cannot, in general, be represented by a simple correction to vortex speed or inertia.
\end{itemize}

\FloatBarrier
\Needspace{0.78\textheight}
\subsection{Recovery of the small-amplitude limit}\label{sec:smallamp}

\begin{wrapfigure}{r}[0pt]{0.56\textwidth}
  \centering
  \captionsetup{font=small}
  \vspace{-0.45\baselineskip}
  \begin{subfigure}[t]{1.16\linewidth}
    \centering
    \paperfigure[width=\linewidth,height=0.36\textheight,keepaspectratio]{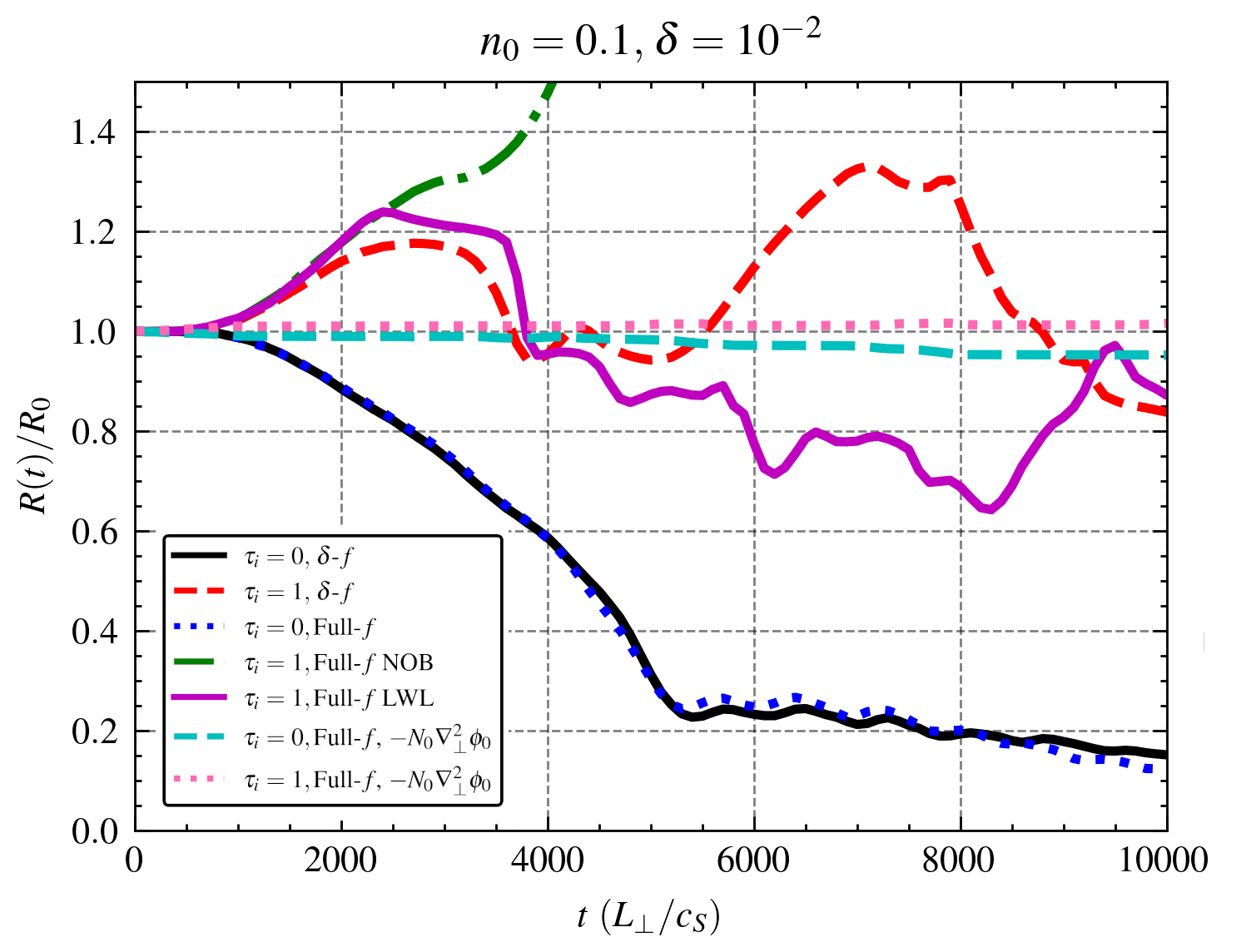}
    \caption{$n_0/N_{\mathrm{mean}}=0.1$, $\Omega_0=0.01$.}
  \end{subfigure}

  \vspace{0.20em}
  \begin{subfigure}[t]{1.16\linewidth}
    \centering
    \paperfigure[width=\linewidth,height=0.36\textheight,keepaspectratio]{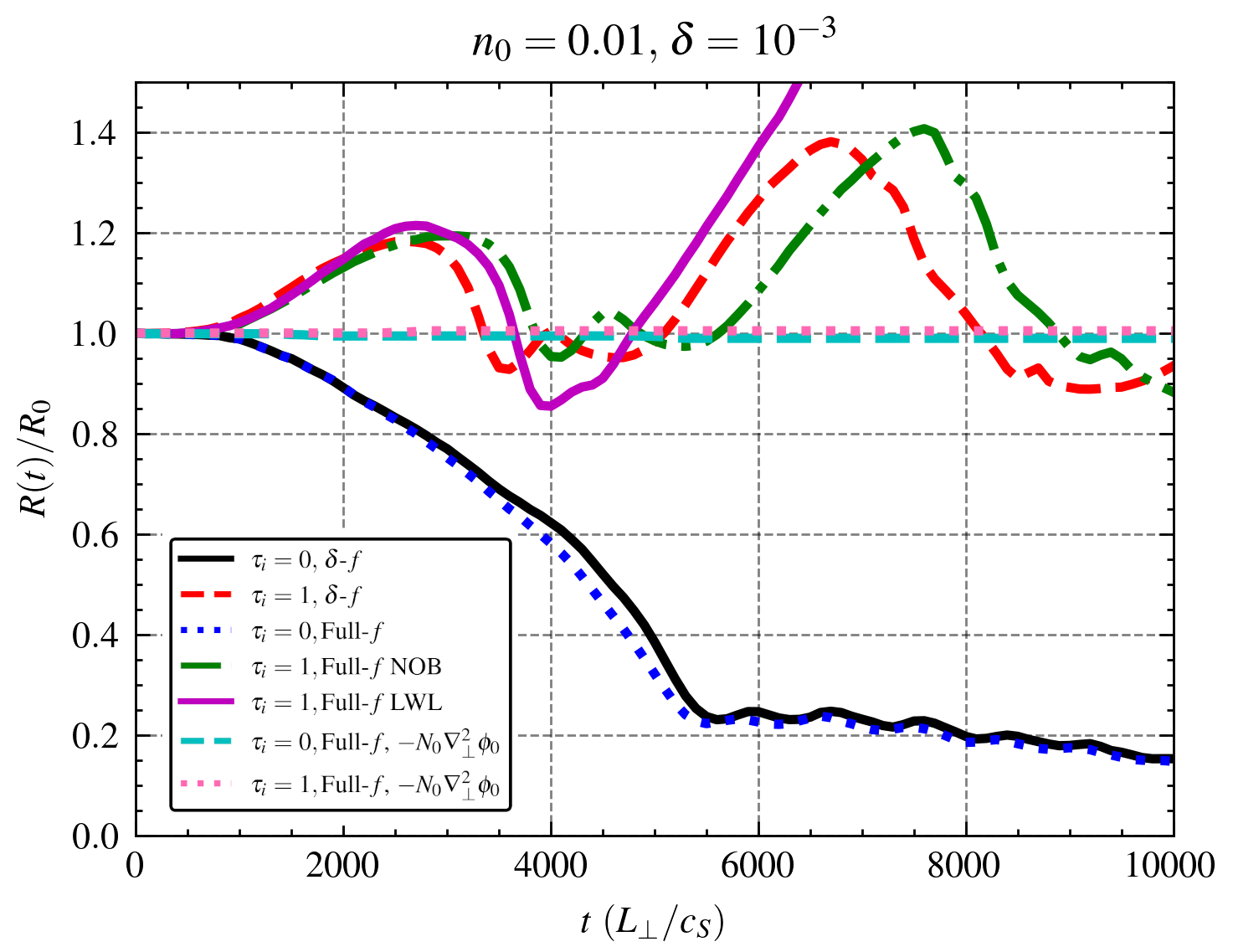}
    \caption{$n_0/N_{\mathrm{mean}}=0.01$, $\Omega_0=0.001$.}
  \end{subfigure}
  \caption{Separation histories for co-rotating positive-vorticity pairs in $\delta f$ and full-$f$, demonstrating recovery of the local limit as the localized density amplitude decreases relative to $N_{\mathrm{mean}}$. Curves with the $N_{\mathrm{mean}}\lapperp\phi_G$ contribution removed isolate the leading-order term responsible for the local $\delta f$ limit.}
  \label{fig:small_amp}
  \vspace{-0.25\baselineskip}
\end{wrapfigure}

A necessary validation of the generalized initialization is that the full-$f$ evolution approaches the $\delta f$ result as the localized density amplitude is reduced relative to $N_{\mathrm{mean}}$. This is shown in \cref{fig:small_amp}. At $n_0/N_{\mathrm{mean}}=0.1$, the cold-ion trajectories are already close while warm-ion cases still develop significant deviations at later time. At $n_0/N_{\mathrm{mean}}=0.01$, with the vorticity amplitude reduced by the same factor, the $\delta f$ and full-$f$ separation curves are substantially more similar. The full-$f$ evolution is slightly delayed, consistent with the additional inertia associated with the finite mean-density contribution to polarization.

The same comparison also identifies the term responsible for the local limit. If the mean-density contribution proportional to $N_{\mathrm{mean}}\lapperp\phi_G$ is removed from the initial full-$f$ polarization balance, the corresponding vortices remain nearly stationary in the small-amplitude cases. This is consistent with an amplitude ordering in which the mean-density term is $\mathcal O(\epsilon)$ while the additional density-dependent polarization terms are $\mathcal O(\epsilon^2)$. Thus the constant-background polarization equation of the $\delta f$ model emerges as the leading-order contribution of the full-$f$ equation, even though $N_{\mathrm{mean}}$ itself is not a static full-$f$ background field.

\FloatBarrier
\subsection{Finite-amplitude mean-density polarization}\label{sec:background_pol}

We next consider a finite-amplitude pair with $n_0=N_{\mathrm{mean}}=1$, $a=6\rho_S$, $R_0=4a$, $|\Omega_0|=0.01$, domain $(128\rho_S)^2$ and $512^2$ grid points. In this regime the localized density excess is comparable to the mean-density level and a full-$f$ treatment is required. The hyperviscosity is $\nu=10^{-5}$, a factor of ten larger than in the corresponding $\delta f$ studies, because substantially finer scales develop in the finite-amplitude calculation.

\begin{figure}[!htbp]
  \centering
  \paperfigure[width=\linewidth]{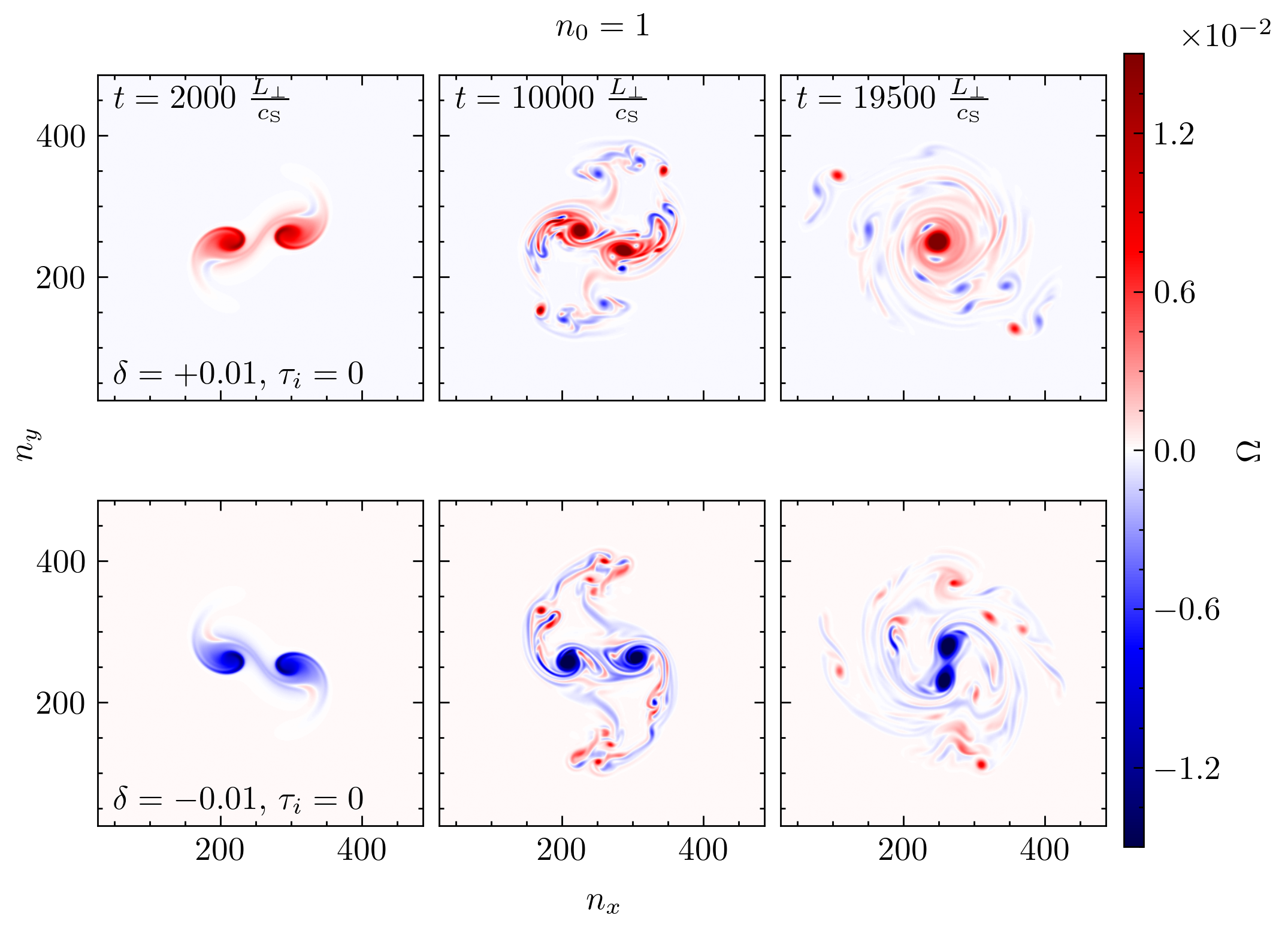}
  \caption{Finite-amplitude full-$f$ cold-ion ($\tau_i=0$) co-rotating vortex evolution for positive and negative initial vorticity. Opposite-sign vorticity is generated around the original cores and the positive--negative symmetry breaks early. The positive-vorticity pair merges at approximately $t=15000\,L_\perp/c_S$, whereas the negative-vorticity pair is still merging at $t=20000\,L_\perp/c_S$. Parameters: $(L_x,L_y)=(128\rho_S,128\rho_S)$, numerical resolution $512^2$, $\Omega_0=\delta=\pm0.01$, $n_0=N_{\mathrm{mean}}=1$, $a=6\rho_S$, $R_0=4a$, and $\nu=10^{-5}$. The plotted $n_x,n_y$ axes denote grid indices.}
  \label{fig:full_cold}
\end{figure}

\begin{figure}[!htbp]
  \centering
  \paperfigure[width=\linewidth]{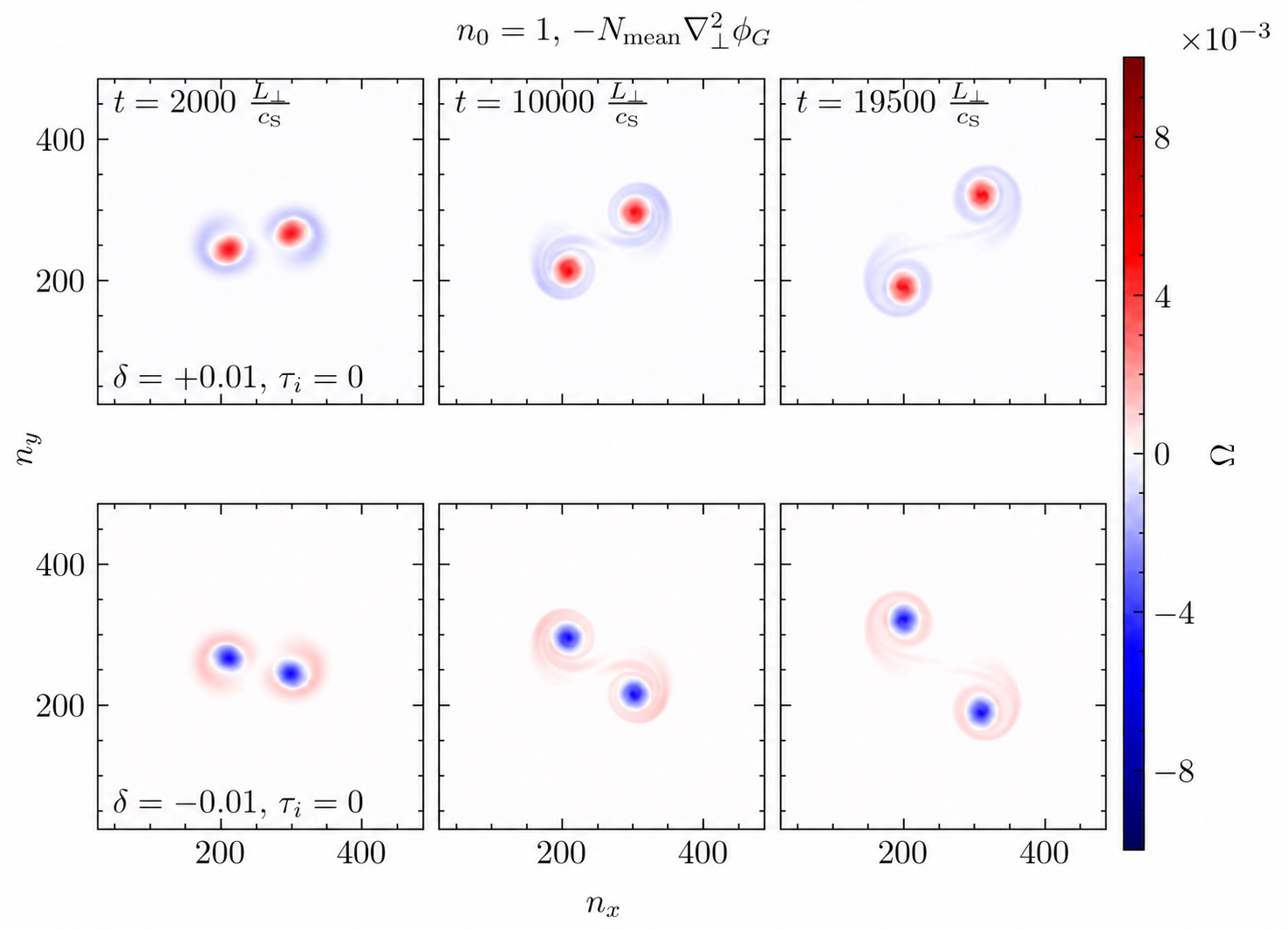}
  \caption{Same finite-amplitude cold-ion initialization as \cref{fig:full_cold}, with the mean-density component $-N_{\mathrm{mean}}\lapperp\phi_G$ subtracted initially. The remaining contribution is the additional vorticity generated by the density-dependent part of the full-$f$ polarization response. The resulting shielded-vorticity structure exhibits positive--negative symmetry and drives early mutual repulsion.}
  \label{fig:background_only}
\end{figure}

For cold ions, the generalized full-$f$ initialization produces a strong departure from the local dynamics, as shown in \cref{fig:full_cold}. Opposite-sign vorticity is generated around the original cores, the positive--negative symmetry breaks early, and merger is delayed. Positive vortices merge only around $t\simeq15000\,L_\perp/c_S$, while the negative pair remains in the merger process at the end of the run, $t=20000\,L_\perp/c_S$. The origin of the delay can be isolated by subtracting the mean-density component $-N_{\mathrm{mean}}\lapperp\phi_G$ from the initial vorticity balance, as shown in \cref{fig:background_only}. What remains is the additional vorticity generated by the density-dependent part of the full-$f$ polarization response. It forms a shielded structure with opposite-sign surrounding vorticity and generates an initially repulsive contribution. The full finite-amplitude dynamics can therefore be interpreted as a competition between the ordinary vortex interaction and a nonlinear mean-density/density-variation polarization response absent from the local model.

\FloatBarrier
\subsection{Density--vorticity symmetry breaking at finite amplitude}\label{sec:full_symmetry_results}

\begin{figure}[!htbp]
  \centering
  \paperfigure[width=0.90\linewidth,height=0.78\textheight,keepaspectratio]{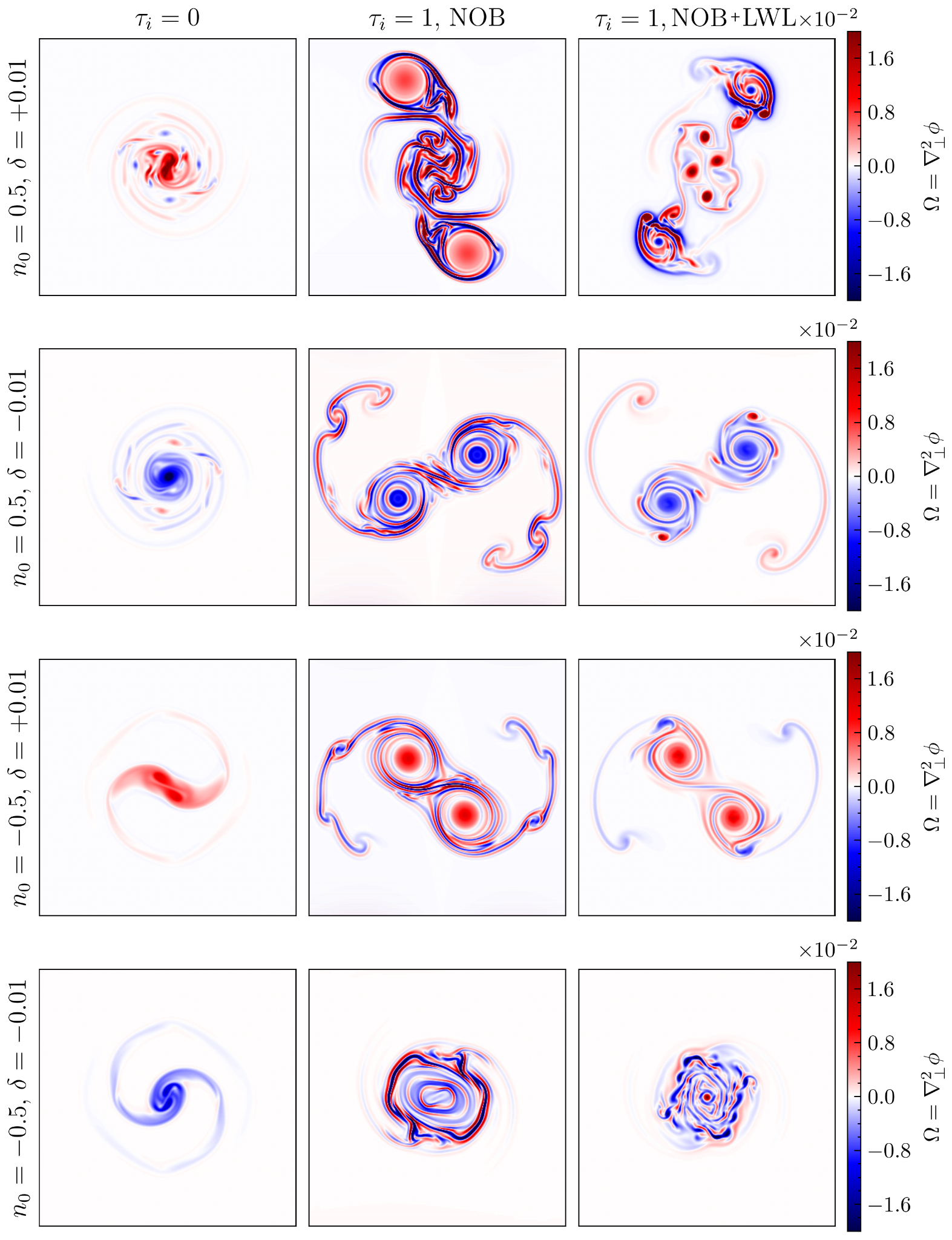}
  \caption{Independent density--vorticity sign scan for finite-amplitude full-$f$ vortex pairs. The four rows correspond to the states $B_+$, $B_-$, $H_+$ and $H_-$ defined in \cref{eq:sym_scan_states}. Comparing $B_\sigma$ with $H_\sigma$ applies $\mathcal S_n$ and reverses the localized density departure at fixed $\Omega_0$; comparing opposite subscripts applies $\mathcal S_\Omega$ and reverses the circulation at fixed density character. Columns show the cold-ion case ($\tau_i=0$), warm-ion arbitrary-wavelength NOB polarization ($\tau_i=1$), and warm-ion NOB+LWL polarization ($\tau_i=1$). The cold-ion column already violates the local $\mathcal S_n$ symmetry through finite-amplitude NOB polarization, whereas the warm-ion columns combine this effect with the FLR symmetry breaking of the local gyrofluid dynamics. The $\times10^{-2}$ annotations belong to the vorticity color scales.}
  \label{fig:inverse_density_scan}
\end{figure}

The generalized full-$f$ initialization allows the density and vorticity signs to be varied independently. With
\begin{equation}
  G(x,y)\equiv g_1(x,y)+g_2(x,y),
  \qquad
  \Delta n_{e,0}=n_0 G,
  \qquad
  \Omega_0=\delta G,
  \label{eq:sym_scan_initial_fields}
\end{equation}
we define two elementary sign operations,
\begin{align}
  \mathcal S_n:&\quad (n_0,\delta)\mapsto(-n_0,\delta),
  & (\Delta n_{e,0},\Omega_0)&\mapsto(-\Delta n_{e,0},\Omega_0),
  \label{eq:sym_scan_Sn}\\
  \mathcal S_\Omega:&\quad (n_0,\delta)\mapsto(n_0,-\delta),
  & (\Delta n_{e,0},\Omega_0)&\mapsto(\Delta n_{e,0},-\Omega_0).
  \label{eq:sym_scan_So}
\end{align}
Their composition $\mathcal S_{n\Omega}=\mathcal S_n\mathcal S_\Omega$ reverses both the localized density departure and the circulation. For the scan shown in \cref{fig:inverse_density_scan}, it is convenient to denote
\begin{equation}
  B_\sigma:\ (n_0,\delta)=(+0.5,\sigma\,0.01),
  \qquad
  H_\sigma:\ (n_0,\delta)=(-0.5,\sigma\,0.01),
  \qquad \sigma=\pm1,
  \label{eq:sym_scan_states}
\end{equation}
where $B$ and $H$ denote density blobs and holes, while the subscript gives the sign of the initial vorticity. Thus
\begin{equation}
  \mathcal S_n:B_\sigma\leftrightarrow H_\sigma,
  \qquad
  \mathcal S_\Omega:B_+\leftrightarrow B_-,\ H_+\leftrightarrow H_-,
  \label{eq:sym_scan_pairs}
\end{equation}
provides two independent tests of density and circulation parity.

In the cold-ion local model, $D_\phi\Omega=0$ [\cref{eq:euler_symmetry}], so $\mathcal S_n$ is invisible to the vorticity dynamics: $B_\sigma$ and $H_\sigma$ have identical $\Omega_0$ and therefore the same Euler-like evolution. At finite $\tau_i$, this local density symmetry is broken at leading FLR order by \cref{eq:df_symmetry_breaking}; under $\mathcal S_n$,
\begin{equation}
  W=\Omega+\frac{\tau_i}{2}\lapperp\widetilde n_e
  \mapsto
  \Omega-\frac{\tau_i}{2}\lapperp\widetilde n_e,
  \qquad
  \frac{\tau_i}{2}\bracket{\Omega}{\widetilde n_e}
  \mapsto
  -\frac{\tau_i}{2}\bracket{\Omega}{\widetilde n_e},
  \label{eq:sym_scan_df_breaking}
\end{equation}
which is the Kendl symmetry breaking of the local warm-ion system \cite{vortexmerging}.

Full-$f$ polarization adds a second mechanism. In the LWL form of the NOB field equation,
\begin{equation}
  \mathcal Q
  =N_{\mathrm{mean}}\Omega
  +\underbrace{\Delta N_i\,\Omega
  +\gradperp\Delta N_i\!\cdot\!\gradperp\phi}_{\mathcal Q_{\mathrm{NOB}}},
  \label{eq:sym_scan_lwl_split}
\end{equation}
only the first term survives the local OB reduction. The nonlinear contribution $\mathcal Q_{\mathrm{NOB}}$ depends explicitly on the reconstructed density departure. Consequently, at fixed prescribed $\Omega_0$, the transformation $\mathcal S_n$ requires a different $N_i$ reconstruction and need not preserve the field inversion. Likewise, under the simultaneous finite-amplitude sign reversal $(\Delta N_i,\phi)\mapsto(-\Delta N_i,-\phi)$, the OB term is odd but $\mathcal Q_{\mathrm{NOB}}$ is even [\cref{eq:full_parity_test,eq:full_lwl_symmetry}]. Hence the full-$f$/NOB system has no exact blob--hole parity even at $\tau_i=0$.

This separation is visible directly in \cref{fig:inverse_density_scan}. In the cold-ion column, the pairs
\begin{equation}
  B_+\not\sim H_+,
  \qquad
  B_-\not\sim H_-,
  \label{eq:sym_scan_cold_result}
\end{equation}
show that reversing the density departure at fixed initial vorticity already produces different finite-amplitude vorticity fields; this difference cannot originate from ion-FLR gyroaveraging. At $\tau_i=1$, the same NOB asymmetry is superposed with the local FLR mechanism of \cref{eq:sym_scan_df_breaking}. The arbitrary-wavelength NOB column therefore contains both symmetry-breaking contributions, whereas the NOB+LWL column removes part of the short-wavelength polarization response while retaining the nonlinear density dependence. The resulting morphologies confirm that the two mechanisms coexist rather than replace one another: full-$f$ polarization breaks density parity through $N_i$ in the field inversion, while finite-$\rho_i$ dynamics remains sensitive to the relative density--vorticity configuration.

The scan may therefore be summarized schematically as
\begin{equation}
  \boxed{\begin{array}{lll}
    \tau_i=0,\ \delta f: & \mathcal S_n\ \text{symmetry},\\[0.15em]
    \tau_i>0,\ \delta f: & \mathcal S_n\ \text{broken by FLR},\\[0.15em]
    \tau_i=0,\ \text{full-}f: & \mathcal S_n\ \text{broken by NOB polarization},\\[0.15em]
    \tau_i>0,\ \text{full-}f: & \text{NOB and FLR symmetry breaking coexist}.
  \end{array}}
  \label{eq:sym_scan_summary}
\end{equation}
Thus the initial vorticity field $\Omega_0(\bm r)$ alone is insufficient to determine a finite-amplitude gyrofluid vortex trajectory: the complete density configuration enters both through the nonlinear polarization inertia and, for warm ions, through the FLR-modified advection.

\FloatBarrier
\subsection{Arbitrary-wavelength (second order Padé) polarization effects}\label{sec:warm_full}

\begin{figure}[!htbp]
  \centering
  \paperfigure[width=\linewidth]{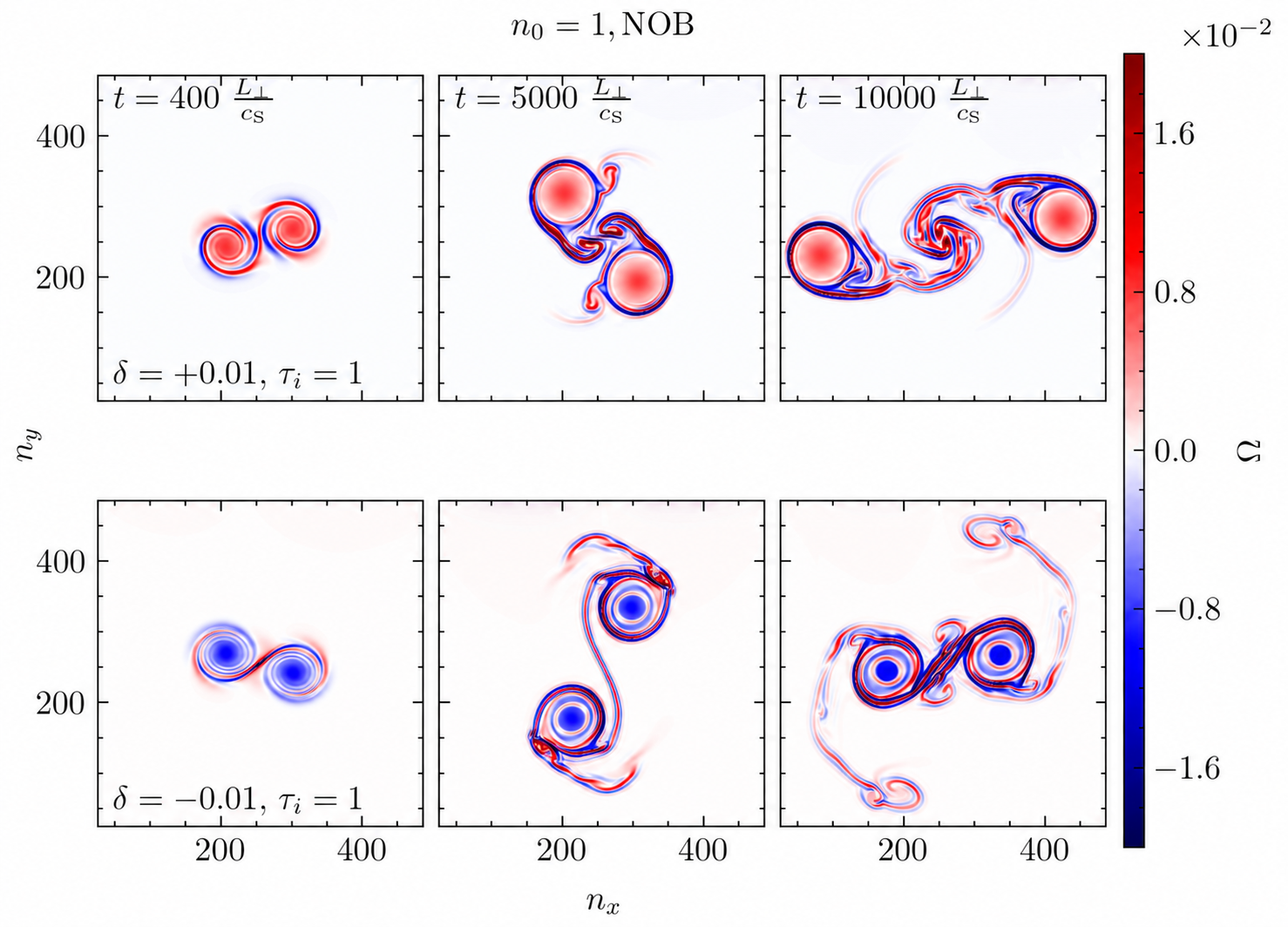}
  \caption{Finite-amplitude full-$f$ warm-ion ($\tau_i=1$) co-rotating vortex evolution with arbitrary-wavelength NOB polarization for positive and negative initial vorticity. Parameters: $(L_x,L_y)=(128\rho_S,128\rho_S)$, numerical resolution $512^2$, $\Omega_0=\delta=\pm0.01$, $n_0=N_{\mathrm{mean}}=1$, $a=6\rho_S$, $R_0=4a$, and $\nu=10^{-5}$. The plotted $n_x,n_y$ axes denote grid indices.}
  \label{fig:full_warm_nob}
\end{figure}

For $\tau_i=1$ the additional mean-density polarization contribution competes with FLR gyrospinning. The full arbitrary-wavelength NOB and NOB+LWL polarizations can then produce markedly different finite-amplitude evolutions, particularly for positive initial vorticity; the corresponding arbitrary-wavelength NOB evolution is shown in \cref{fig:full_warm_nob}. In the NOB case, finely structured enveloping vorticity sheets are retained and the vortices orbit at nearly constant separation up to approximately $t=1750\,L_\perp/c_S$. The separation subsequently decreases slightly before a rapid mutual repulsion at $t\simeq3750\,L_\perp/c_S$. The initial orbit indicates that the attractive component associated with FLR spin-up can temporarily balance the repulsive mean-density polarization contribution.

\begin{figure}[!htbp]
  \centering
  \paperfigure[width=\linewidth]{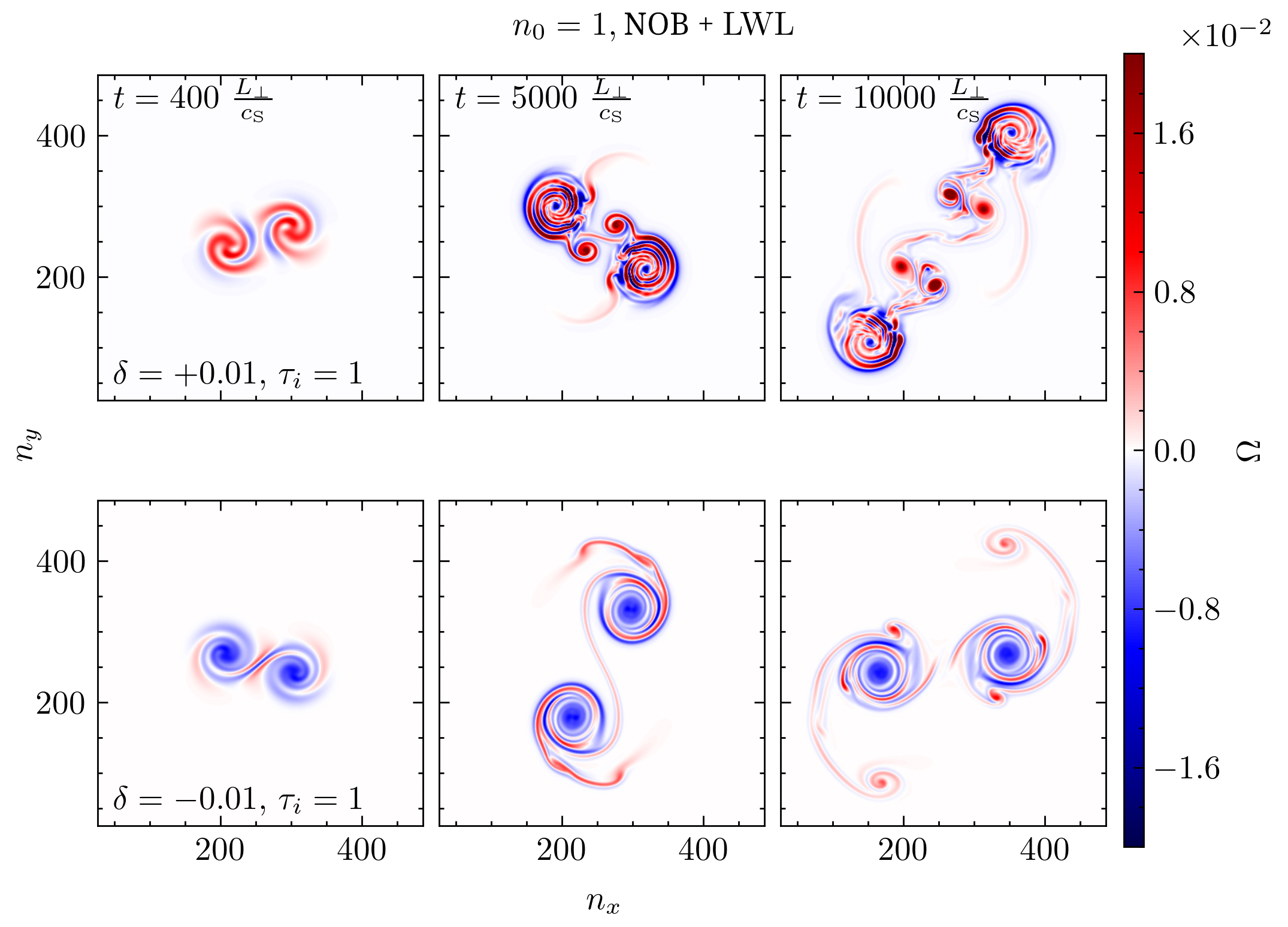}
  \caption{Same warm-ion finite-amplitude co-rotating setup as \cref{fig:full_warm_nob}, evolved with NOB+LWL polarization. For the positive-vorticity branch, the LWL model develops coarser structures and pronounced vortex-shedding behavior, which is not observed in the arbitrary-wavelength NOB evolution over the simulated interval; the negative-vorticity NOB and NOB+LWL evolutions are more similar.}
  \label{fig:full_warm}
\end{figure}

\begin{figure}[!htbp]
  \centering
  \paperfigure[width=0.95\linewidth]{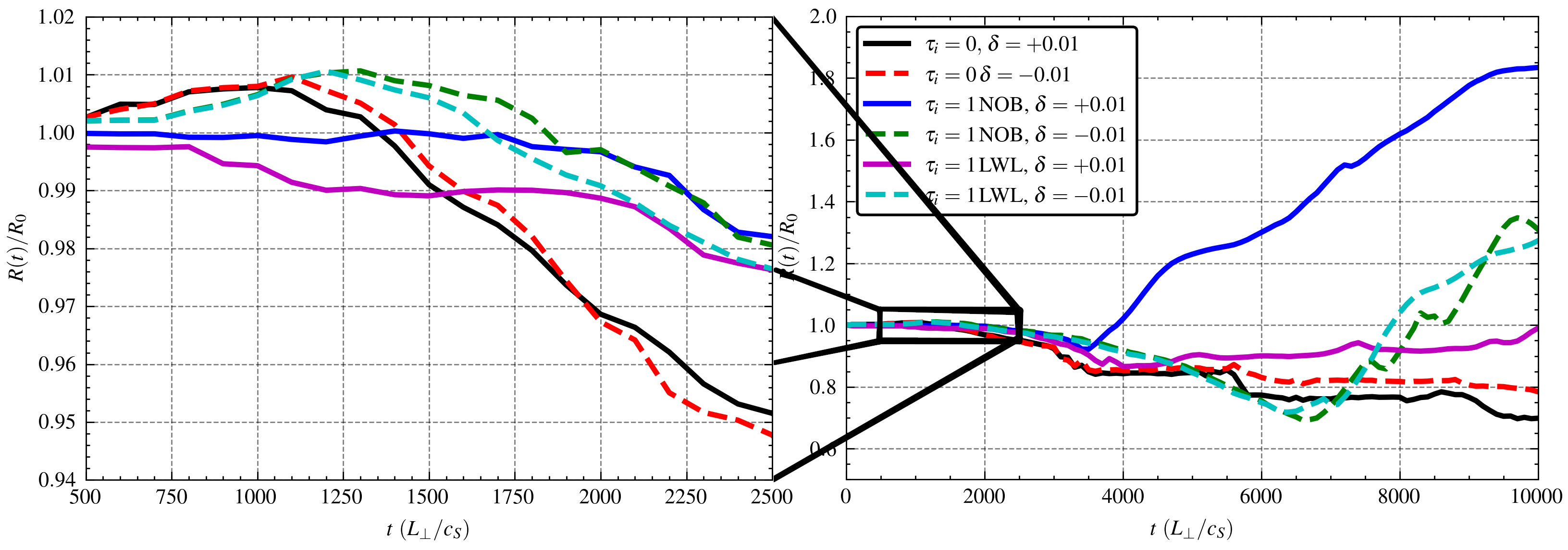}
  \caption{Separation histories for the generalized finite-amplitude full-$f$ co-rotating simulations. The warm-ion NOB case exhibits an extended nearly constant-separation orbit followed by rapid repulsion, illustrating the competition between FLR-induced attraction and the repulsive mean-density polarization component.}
  \label{fig:full_distances}
\end{figure}

The NOB+LWL case, shown in \cref{fig:full_warm}, suppresses part of the short-scale polarization structure. In the simulations this results in coarser evolved structures and pronounced vortex-shedding behavior reminiscent of a K\'arm\'an-type street; no corresponding shedding is observed in the arbitrary-wavelength NOB evolution over the simulated interval. For negative initial vorticity, by contrast, the NOB and NOB+LWL trajectories are more similar, consistent with the repelling FLR branch already dominating the interaction. The strong sign dependence emphasizes that the finite-amplitude correction is not a uniform change in inertia or interaction strength: it changes the spatial vorticity structure that then interacts nonlinearly with the FLR-generated sheets. The corresponding separation histories are collected in \cref{fig:full_distances}.

\FloatBarrier
\section{Discussion}\label{sec:discussion}

The $\delta f$ and full-$f$ calculations expose two distinct ways in which gyrofluid polarization changes an otherwise familiar two-dimensional vortex-pair problem. Within the local model, finite ion temperature modifies the response kernel and generates gyrospinning, thin vorticity sheets and a separation-dependent orbital response. The analytical model isolates this effect by replacing the Laplacian Green function with \cref{eq:Gk}. Although the closed form was obtained from a reduced polarization relation, the good direct comparison with the LWL orbital frequencies indicates that the kernel picture captures a meaningful part of the interaction physics. The observed dip and the arrested-merger regime should nevertheless not be conflated: the former is a prediction of the reduced LWL interaction model, whereas the latter is a nonlinear morphological transition in the full arbitrary-wavelength $\delta f$ simulations.

The full-$f$ extension introduces an additional mechanism. Because the complete $N_i$ multiplies the polarization response, the finite mean-density level and the localized density variation both participate in the field inversion. The mean-density contribution together with the nonlinear density-dependent terms produces vorticity outside the prescribed Gaussian cores. In the cold-ion case this additional contribution is directly visible as a shielded, initially repulsive structure. At finite $\tau_i$ it competes with the FLR-generated vorticity sheets. This explains why the difference between local and global dynamics becomes qualitative once $n_0/N_{\mathrm{mean}}=\mathcal O(1)$, while the same generalized initialization correctly approaches the $\delta f$ result as $n_0/N_{\mathrm{mean}}\rightarrow0$.

The independent sign scan in \cref{fig:inverse_density_scan} makes the associated symmetry structure explicit. In the cold-ion local model, density does not enter the Euler vorticity equation, whereas the warm-ion $\delta f$ system loses this density--vorticity equivalence through FLR polarization, as identified by Kendl \cite{vortexmerging}. In full-$f$, a second symmetry breaking appears already at $\tau_i=0$ because the spatially varying $N_i$ multiplies the polarization response. At finite $\tau_i$ the two mechanisms coexist rather than replacing one another: the nonlinear NOB field inversion distinguishes density enhancements from depletions, while gyroaveraging remains sensitive to the relative density--vorticity configuration. The absence of a simple sign-reversed partner in the warm-ion scan is therefore a direct consequence of these two parity-breaking contributions acting simultaneously.

The comparison between arbitrary-wavelength NOB and NOB+LWL polarization also suggests caution when using a long-wavelength closure in strongly filamenting vortex dynamics. In the present finite-amplitude positive-vorticity cases, suppressing the shorter-scale polarization response changes the morphology, stability and shedding of the vortices. The result does not imply that an LWL treatment is generally invalid; rather, it identifies a regime in which the fine vorticity sheets generated by FLR and finite-amplitude polarization participate directly in the subsequent interaction and therefore cannot be assumed dynamically passive.

\FloatBarrier
\section{Conclusions}\label{sec:conclusion}

We have used a deliberately elementary vortex-pair configuration to separate two effects that are often entangled in more complete plasma-turbulence simulations: FLR modification of the gyrofluid response and finite-amplitude modification of the polarization itself. The main results are:

\begin{enumerate}[label=\textbf{R\arabic*.},leftmargin=2.2em,itemsep=0.55em,topsep=0.45em]
  \item \textbf{FLR physics can be expressed as an effective interaction kernel.} The reduced $\delta f$ polarization relation gives a finite-core orbital-frequency law in which the short-range Euler interaction is regularized on the FLR scale while the far field remains Euler-like. Direct $\delta f$ LWL simulations reproduce the predicted $\tau_i$-dependent dip in the orbital frequency.

  \item \textbf{Increasing ion temperature produces a distinct, separation-dependent merger--orbit transition.} Positive-vorticity $\delta f$ pairs change from complete merger to long-lived co-rotating states as $\tau_i$ is increased, but the crossover value depends on the initial vortex separation and is not a universal critical temperature ratio. Within each scan, the onset of long-lived co-rotation coincides with a pronounced increase of enstrophy in the fine, FLR-generated vorticity sheets, indicating that these sheets become an active part of the interaction rather than a passive by-product of gyrospinning.

  \item \textbf{The local $\delta f$ model is recovered as a controlled small-amplitude limit of the generalized full-$f$ initialization.} The mean-density contribution to the full-$f$ polarization is the leading $\mathcal O(\epsilon)$ term and yields the constant-background $\delta f$ polarization at leading order, while the additional density-dependent nonlinear polarization terms enter at higher order. Accordingly, the full-$f$ and $\delta f$ trajectories approach one another as the matched localized density amplitude is reduced.

  \item \textbf{Finite amplitude introduces a second, genuinely full-$f$ symmetry breaking.} Writing $\Delta n_{e,0}=n_0G$ and $\Omega_0=\delta G$ makes density reversal $(n_0,\delta)\mapsto(-n_0,\delta)$ and circulation reversal $(n_0,\delta)\mapsto(n_0,-\delta)$ independent operations. In the cold-ion local model the first is a symmetry because $D_\phi\Omega=0$; in full-$f$ it is broken already at $\tau_i=0$ by the nonlinear NOB terms $\Delta N_i\,\Omega+\gradperp\Delta N_i\cdot\gradperp\phi$. Warm ions then superpose the local FLR symmetry breaking $\propto\tau_i\bracket{\Omega}{\widetilde n_e}$ on this finite-amplitude NOB asymmetry.

  \item \textbf{Arbitrary-wavelength response controls the morphology of the symmetry-broken warm-ion state.} When the NOB and FLR mechanisms act simultaneously, retaining arbitrary-wavelength polarization preserves finer vorticity sheets and can produce trajectories that differ qualitatively from NOB+LWL evolution, including different repulsion and shedding behavior.
\end{enumerate}

The central implication is that the passage from $\delta f$ to full-$f$ cannot in general be represented by a simple correction to vortex speed or inertia. Once the localized full-$f$ density variation is comparable to the mean-density level, the nonlinear polarization equation changes the spatial vorticity field that mediates the interaction and removes the simple density--vorticity sign equivalence of the local cold-ion problem. In the controlled two-vortex system studied here, the combined NOB and FLR symmetry breaking is already sufficient to determine whether vortices merge, remain in a long-lived orbit, separate or shed secondary structures. These configurations therefore provide sensitive benchmark problems for future full-$f$ gyrofluid developments and a natural starting point for testing the same mechanisms in driven vortex ensembles and edge-turbulence models.

\section*{Data availability statement}
The GREENY source code used for the simulations is available at \url{https://git.uibk.ac.at/c7441315/greeny}. The data that support the findings of this study are available from the corresponding author upon reasonable request.

\section*{Acknowledgements}

F.F.L. acknowledges financial support by KKKÖ of ÖAW.
M.H. acknowledges support by the UiT Aurora Centre Program, UiT The Arctic University of Norway (2020).
This research was funded in whole or in part by the Austrian Science Fund (FWF) [P 34241-N].
The authors thank Lukas Einkemmer for insightful discussions and valuable advice concerning the numerical inversion of the polarization equation.

\FloatBarrier
\section*{Author contributions}
\textbf{M. Rinner}: Data curation (lead); Conceptualization (equal); Validation (equal); Software (equal); Visualization (lead); Writing -- original draft (equal); Methodology (equal); Formal analysis (lead).
\textbf{F. F. Locker}: Supervision (equal); Conceptualization (equal); Writing -- original draft (equal); Software (equal); Methodology (equal); Visualization (support); Formal analysis (support).
\textbf{M. Held}: Supervision (equal); Conceptualization (equal).
\textbf{A. Kendl}: Supervision (equal); Resources (lead); Project administration (lead).

\bibliographystyle{unsrtnat}
\bibliography{references}

\end{document}